\documentclass[usegraphicx,usenatbib]{mn2e}
\usepackage{color}
\input{colordvi.tex}

\newcommand{\ltilde} {~ \raisebox{-1ex}{$\stackrel{\textstyle <}{\sim}$} ~}
\begin{document}
\title[Limits on Isocurvature Non-Gaussianity]{Limits on
Isocurvature Perturbations from Non-Gaussianity in WMAP Temperature Anisotropy}
\author[Hikage et al.]{Chiaki~Hikage$^1$\thanks{hikage@astro.princeton.edu},
  Kazuya~Koyama$^2$, Takahiko~Matsubara$^3$, Tomo~Takahashi$^4$,
\newauthor Masahide~Yamaguchi$^5$ \\
\\
$^1$
  Department of Astrophysical Sciences, Princeton University, Peyton Hall,
  Princeton NJ 08544, USA \\
$^2$
  Institute of Cosmology and Gravitation, University of Portsmouth,
  Portsmouth PO1 2EG \\
$^3$
  Department of Physics and Astrophysics,
  Nagoya University, Chikusa, Nagoya 464-8602, Japan \\
$^4$
  Department of Physics, Saga University, Saga 840-8502, Japan \\
$^5$
  Department of Physics and Mathematics, Aoyama Gakuin University,
  Sagamihara 229-8558, Japan and\\ \,\,\,
  Department of Physics, Stanford University, Stanford CA 94305, USA
}
\maketitle
\begin{abstract}
We study the effect of primordial isocurvature perturbations on
non-Gaussian properties of CMB temperature anisotropies. We consider
generic forms of the non-linearity of isocurvature perturbations which
can be applied to a wide range of theoretical models.  We derive
analytical expressions for the bispectrum and the Minkowski Functionals
for CMB temperature fluctuations to describe the non-Gaussianity from
isocurvature perturbations.  We find that the isocurvature
non-Gaussianity in the quadratic isocurvature model, where the
isocurvature perturbation $S$ is written as a quadratic function of
the Gaussian variable $\sigma$, $S=\sigma^2-\langle \sigma^2\rangle$,
can give the same signal-to-noise as $f_{\rm NL}=30$ even if we impose
the current observational limit on the fraction of isocurvature
perturbations contained in the primordial power spectrum
$\alpha$.  We give constraints on isocurvature
non-Gaussianity from Minkowski Functionals using the WMAP 5-year data.  We
do not find a significant signal of isocurvature
non-Gaussianity. For the quadratic isocurvature model, we obtain a
stringent upper limit on the isocurvature fraction $\alpha<0.070$
(95\% CL) for a scale invariant spectrum which is comparable to the
limit obtained from the power spectrum.
\end{abstract}

\section{Introduction}
\label{sec:intro}

Recent observational progress in cosmology represented by surveys 
such as WMAP and SDSS has enabled a detailed analysis of cosmic density fields to
investigate the physics of the early universe.  In particular,
(non-)Gaussianity of primordial density fields has been received much
attention recently as a key observational probe to differentiate
models of the early universe.  In the simplest single field inflationary
scenario, quantum fluctuations of the inflaton during inflation are
assumed to be the origin of cosmic density fluctuations and such a model
predicts adiabatic and almost Gaussian primordial fluctuations.
However, other generation mechanisms of density
fluctuations such as the curvaton scenario
\citep{Mollerach1990,LM1997,MT2001,ES2002,LW2002}, modulated
reheating \citep{Kofman2003,DGZ2004} and so on have been
proposed. In these mechanism, the nature of primordial density
fluctuations can be very different from that of a simple inflation
model, in particular, in terms of Gaussianity of the fluctuations.  In
fact, it has been shown that large non-Gaussianity can be generated in
curvaton models
\citep{Lyth2003,BMR2004,EN2005,ML2006,SVW2006,AVW2007,Huang2008,ISTY2008a,ET2008},
modulated reheating scenarios \citep{Zaldarriaga2004,SY2008, ISTY2008b},
Dirac-Born-Infeld inflation models \citep{AST2004,CHKS2007,LRST2008,AMK2008},
Ghost inflation \citep{ACMZ2004}, ekpyrotic models
\citep{KMVW2007,BKO2008,LS2008}, single field inflation with a
feature in its potential \citep{CEL2007}, a Gaussian-squared component
in curvature or entropy perturbations \citep{LM1997,BL2006,ST2008}
and multi-brid inflation \citep{Sasaki2008,NS2009}.

The primordial non-Gaussianity has been quantitatively measured by
higher-order statistics such as bispectra and Minkowski
Functionals from CMB temperature maps obtained by WMAP
\citep{Komatsu2008,Creminelli2007,YW2008,Hikage2008} and also from
large-scale structure \citep{Slosar2008}. So far the observational
results are consistent with the Gaussian hypothesis.  There is,
however, a hint of primordial non-Gaussianity at 2-3 $\sigma$ level
\citep{YW2008,Komatsu2008}. Since the non-Gaussianity predicted in the
simplest inflation model is too small to be
detected by current observations, if a non-Gaussian signal is observed
and it originates from primordial fluctuations, other mechanisms beyond
the simplest model would be required in the dynamics of early
universe.

As another probe of the early universe, the adiabaticity of primordial
density fields has also been the subject of intense study.  In fact,
current cosmological observations of TT and TE spectra of CMB with
some other distance measurements such as type Ia supernovae (SNe) and
baryon acoustic oscillation (BAO) have already constrained the
fraction of isocurvature perturbations to be less than 10\%
\citep[e.g.,][]{BDP2006,KS2007,Komatsu2008}. Examples of
isocurvature perturbations along with non-Gaussianity have been
discussed in the context of non-Gaussian field potentials
\citep{LM1997,Peebles1999,BL2006,ST2008}, the curvaton scenario
\citep{Lyth2003,BMR2004,Beltran2008,MT2008}, modulated reheating
\citep{BC2006}, baryon asymmetry \citep{KNT2009} and the axion
\citep{KNSST2008,KNSST2009}.  In particular, non-Gaussianity
generated from isocurvature perturbations has been systematically
investigated recently \citep{KNSST2008,LVW2008,KNSST2009}.

In this paper we discuss non-Gaussianity generated from the
non-linearity of isocurvature perturbations and study a constraint on non-Gaussianity
from isocurvature perturbations using Minkowski Functionals.  For this
purpose, we derive theoretical expressions for bispectra and Minkowski
Functionals to characterize non-Gaussianity in CMB temperature maps
generated from primordial mixed perturbations of adiabatic and
isocurvature components. We characterize the non-linearity of
isocurvature perturbations in two different forms which are
theoretically motivated; one is a Gaussian variable plus its quadratic
correction (Linear Model).  The other form is given as a quadratic of
Gaussian variables without a liner term (Quadratic Model).  These two
generic forms are applicable to a wide range of isocurvature models
listed above.  Then we give actual limits on the isocurvature
non-Gaussianity using the WMAP 5-year data. As far as we know, this is the
first attempt to put a limit on isocurvature perturbations from the
non-Gaussianity of CMB anisotropies.

In this paper, we focus on CDM isocurvature perturbations. However, it is
straightforward to apply our method to other types of
isocurvature perturbations including baryon and neutrino ones.  We
adopt a set of cosmological parameters at the maximum likelihood values for a
power-law $\Lambda$CDM model obtained from the WMAP 5-year data only fit
\citep{Dunkley2008}; $\Omega_{\rm b}=0.0432$, $\Omega_{\rm
cdm}=0.206$, $\Omega_\Lambda=0.7508$, $H_0=72.4~{\rm
km~s^{-1}~Mpc^{-1}}$, $\tau=0.089$, and $n_\phi=0.961$.  The total
amplitude of primordial power spectra is set to be
$\Delta_{\rm tot}(k=0.002{\rm Mpc}^{-1})=2.41\times 10^{-9}$.

This paper is organized as follows; in \S \ref{sec:pb} we give
two different forms of the non-linear isocurvature perturbations
called the ``Linear Model'' and the ``Quadratic Model''. In \S
\ref{sec:linear} we derive analytical expressions for bispectra to
describe isocurvature non-Gaussianity in CMB temperature
anisotropies in the Linear Model.  The isocurvature non-Gaussianity in 
the Quadratic Model is described in \S \ref{sec:quadratic}.  In
\S \ref{sec:mf}, we present generic perturbative formulae of
Minkowski Functionals that can be applied to CMB temperature maps with
adiabatic and
isocurvature non-Gaussianity. In \S \ref{sec:wmaplimit}, we give
limits on isocurvature non-Gaussianity from the WMAP 5-year
temperature maps using Minkowski Functionals. In \S
\ref{sec:implications}, we discuss implications of our results for an
axion isocurvature model.  \S \ref{sec:summary} is
devoted to a summary and our conclusions.


\section{Non-Linear Adiabatic and Isocurvature Perturbations}
\label{sec:pb}

We consider the admixture of an adiabatic perturbation $\zeta$ with a
CDM isocurvature perturbation ${\cal S}$.  The curvature perturbation is
written up to second order in a local form as
\begin{equation}
\label{eq:adi}
\zeta=\phi+\frac{3}{5}f_{\rm NL}(\phi^2-\langle\phi^2\rangle),
\end{equation}
where $\phi$ is the linear term of $\zeta$ that obeys a Gaussian
statistics. The non-linear parameter $f_{\rm NL}$ represents the
quadratic amplitude of the curvature perturbation $\Phi$ during the 
matter era \citep{KS2001}, which is related to $\zeta$ by $\Phi=(3/5)\zeta$.

An isocurvature perturbation ${\cal S}$ between matter and radiation is
defined as
\begin{equation}
\label{eq:defiso}
{\cal S}\equiv\frac{\delta\rho_m}{\rho_m}
-\frac{3\delta\rho_\gamma}{4\rho_\gamma},
\end{equation}
where $\rho_m$ is the matter energy density and $\rho_\gamma$ is
radiation energy density.  In this paper we focus on a CDM
isocurvature perturbation.

We consider the non-linearity of the isocurvature perturbation in two
different forms.  One is a local form similar to the equation
(\ref{eq:adi}) which has a linear (Gaussian) term with a quadratic
correction
\begin{equation}
\label{eq:iso1}
{\rm I.~~Linear~Model:~~~}{\cal S}=\eta+f_{\rm NL}^{\rm
(ISO)}(\eta^2-\langle\eta^2\rangle),
\end{equation}
where $\eta$ is a Gaussian variable and its non-linearity is characterized
by $f_{\rm NL}^{\rm (ISO)}$.  The isocurvature non-Gaussianity in
this form was studied in the context of the axion \citep{KNSST2008}
and the curvaton scenario \citep{LVW2008}.

The other is the case where the linear term is negligible compared with
the quadratic term \citep[e.g.,][]{LM1997,Peebles1999,BL2006,KNSST2008};
\begin{equation}
\label{eq:iso2}
{\rm II.~~Quadratic~Model:~~~}{\cal S}=\sigma^2-\langle\sigma^2\rangle,
\end{equation}
where $\sigma$ obeys Gaussian statistics.  \citet{LM1997} proposed a
scenario to generate the quadratic form of isocurvature perturbations
by introducing a massive free scalar field oscillating around the
vacuum state which has a zero value. In this scenario, the
isocurvature fluctuation has a blue spectrum and thus we here
consider a wide range for the spectral index ranging from 1 to 3 in
the Quadratic Model. \citet{BL2006} showed that a Gaussian-squared
component of the primordial curvature perturbation would be bounded at 10\%
level by the WMAP bound on the bispectrum.

The auto and cross-correlation power spectra for fluctuations of $I$ and $J$
are defined as
\begin{equation}
\label{eq:defpk}
\langle I_{\mathbf k}J_{\mathbf k^\prime}\rangle
=(2\pi)^3\delta_D^{(3)}({\mathbf k}+{\mathbf k^\prime})P_{IJ}(k),
\end{equation}
and then its dimensionless power is given by
\begin{equation}
\label{eq:defnpk}
\Delta_{IJ}(k)=\frac{P_{IJ}(k)k^3}{2\pi^2},
\end{equation}
where $I$ and $J$ denote $\zeta$ or ${\cal S}$.

We assume the following power-law form of auto and cross power spectra
for the Gaussian variables $\phi$
(eq.[\ref{eq:adi}]), $\eta$ (eq.[\ref{eq:iso1}]), and $\sigma$
(eq.[\ref{eq:iso2}]);
\begin{eqnarray}
\label{eq:plaw}
\Delta_{XX}(k)&=&
A_{X}\left(\frac{k}{k_0}\right)^{n_X-1}, \\
\label{eq:plawcor}
\Delta_{XY}(k)&=&(A_{X}A_{Y})^{1/2}\cos\theta_{XY}
\left(\frac{k}{k_0}\right)^{(n_X+n_Y)/2-1},
\end{eqnarray}
where $X$ and $Y$ denote $\phi$, $\eta$ or $\sigma$.  As $f_{\rm
NL}A_\phi^{1/2}$ is observationally limited to be much smaller than
unity, the power spectrum of the primordial adiabatic perturbation is
given by
\begin{equation}
\label{eq:pkadi}
\Delta_{\zeta\zeta}(k) \simeq \Delta_{\phi\phi}(k).
\end{equation}

We define the fraction of isocurvature perturbation as
\begin{equation}
\label{eq:alpha}
\alpha\equiv\frac{P_{\cal SS}(k_0)}{P_{\zeta\zeta}(k_0)+P_{\cal SS}(k_0)},
\end{equation}
which is the same definition as \citet{BDP2006} for example.  We here
set $k_0=0.002$Mpc$^{-1}$.  The parameter $\alpha$ is related to
another common parameter of adiabaticity $\delta_{\rm
adi}^{(c,\gamma)}$ \citep[e.g., eq. (41) of ][]{Komatsu2008} as
$\delta_{\rm adi}^{(c,\gamma)}=[\alpha/(1-\alpha)]^{1/2}/3$.  The
upper limits of $\alpha$ from WMAP, BAO and SN combined are given by
$0.067$ (95\% CL) for axion-type ($\cos\theta_{\zeta{\cal S}}=0$) and
$0.0037$ (95\% CL) for curvaton-type ($\cos\theta_{\zeta{\cal S}}=-1$)
isocurvature perturbations \citep{Komatsu2008}.

\section{Non-Gaussianity of Isocurvature Perturbations I:  Linear Model}
\label{sec:linear}

\subsection{Initial Perturbation}

In the Linear Model (eq.[\ref{eq:iso1}]), power spectra for
the isocurvature perturbation and its cross term with the
adiabatic perturbation become
\begin{eqnarray}
\label{eq:pkcor}
\Delta_{\cal SS}(k)&\simeq &\Delta_{\eta\eta}(k), \\
\label{eq:pkiso1}
\Delta_{\zeta{\cal S}}(k)&\simeq &\Delta_{\phi\eta}(k).
\end{eqnarray}
Here, we have used the fact that $f_{\rm NL}^{\rm (ISO)} \ltilde
1/\sqrt{\Delta_{\cal SS}} \simeq 1/\sqrt{\alpha \Delta_{\phi\phi}} \sim
10^{6}$. In the case that $f_{\rm NL}^{\rm (ISO)}$ is larger than this
upper limit, the linear term of $\eta$ in Eq.~(\ref{eq:iso1}) is
negligible so that the model should be described by the quadratic
model.

The ratio of the amplitude of the power spectra between $\phi$ and
$\eta$ is written in terms of $\alpha$ (eq.[\ref{eq:alpha}]) as
\begin{equation}
\label{eq:aratio1}
\frac{A_\eta}{A_\phi}=\frac{\alpha}{1-\alpha}.
\end{equation}

We define the bispectra of $\zeta$, ${\cal S}$ and their mixed
contribution as
\begin{eqnarray}
\label{eq:defbis}
\langle I_{\mathbf k_1} J_{\mathbf k_2} K_{\mathbf k_3}\rangle
&\equiv &(2\pi)^3 \delta_D^{(3)}({\mathbf k_1}+{\mathbf k_2}+{\mathbf k_3})
\nonumber \\
&&\times B_{IJK}(k_1,k_2,k_3),
\end{eqnarray}
where $I, J$, and $K$ denote $\zeta$ or ${\cal S}$.

When $\phi$ and $\eta$ are initially uncorrelated,
the adiabatic and isocurvature modes evolve independently
under the linear approximation. There exist bispectra only from each mode
given as
\begin{eqnarray}
\label{eq:pbis_aaa}
B_{\zeta\zeta\zeta}(k_1,k_2,k_3)&=&\frac{6}{5}
f_{NL}[P_{\phi\phi}(k_1)P_{\phi\phi}(k_2) \nonumber \\
&&+P_{\phi\phi}(k_2)P_{\phi\phi}(k_3)
+P_{\phi\phi}(k_3)P_{\phi\phi}(k_1)], \nonumber \\ && \\
\label{eq:pbis_iii1}
B_{\cal SSS}(k_1,k_2,k_3)&=&
2f_{\rm NL}^{\rm (ISO)}
[P_{\eta\eta}(k_1)P_{\eta\eta}(k_2) \nonumber \\
&&+P_{\eta\eta}(k_2)P_{\eta\eta}(k_3)
+P_{\eta\eta}(k_3)P_{\eta\eta}(k_1)]. \nonumber \\ &&
\end{eqnarray}
When $\phi$ and $\eta$ are initially
correlated, the following cross-correlation terms may become important
\begin{eqnarray}
\label{eq:pbis_aai1}
B_{(\zeta\zeta{\cal S})}(k_1,k_2,k_3)
&\equiv & B_{\zeta\zeta{\cal S}}(k_1,k_2,k_3)
+ B_{\zeta{\cal S}\zeta}(k_1,k_2,k_3) \nonumber \\
&&+ B_{{\cal S}\zeta\zeta}(k_1,k_2,k_3) \nonumber \\
&=& 2 f_{\rm NL}^{\rm (ISO)}
[P_{\phi\eta}(k_1)P_{\phi\eta}(k_2) \nonumber \\
&&+P_{\phi\eta}(k_2)P_{\phi\eta}(k_3)
+P_{\phi\eta}(k_3)P_{\phi\eta}(k_1)] \nonumber \\
&&+\frac65 f_{\rm NL} [P_{\phi\phi}(k_1)
\left\{P_{\phi\eta}(k_2)+P_{\phi\eta}(k_3)\right\} \nonumber \\
&&+P_{\phi\phi}(k_2)
\left\{P_{\phi\eta}(k_3)+P_{\phi\eta}(k_1)\right\} \nonumber \\
&&+P_{\phi\phi}(k_3)
\left\{P_{\phi\eta}(k_1)+P_{\phi\eta}(k_2)\right\}
], \\
\label{eq:pbis_aii1}
B_{(\zeta{\cal SS})}(k_1,k_2,k_3)
&\equiv & B_{\zeta{\cal SS}}(k_1,k_2,k_3)
+ B_{{\cal S}\zeta{\cal S}}(k_1,k_2,k_3) \nonumber \\
&&+ B_{{\cal SS}\zeta}(k_1,k_2,k_3) \nonumber \\
&=& \frac{6}{5}f_{\rm NL}
[P_{\phi\eta}(k_1)P_{\phi\eta}(k_2) \nonumber \\
&&+P_{\phi\eta}(k_2)P_{\phi\eta}(k_3)
+P_{\phi\eta}(k_3)P_{\phi\eta}(k_1)] \nonumber \\
&&+ 2 f_{\rm NL}^{\rm (ISO)} [P_{\phi\eta}(k_1)
\left\{P_{\eta\eta}(k_2)+P_{\eta\eta}(k_3)\right\} \nonumber \\
&&+ P_{\phi\eta}(k_2)
\left\{P_{\eta\eta}(k_3)+P_{\eta\eta}(k_1)\right\} \nonumber \\
&&+ P_{\phi\eta}(k_3)
\left\{P_{\eta\eta}(k_1)+P_{\eta\eta}(k_2)\right\}].
\end{eqnarray}

\subsection{Bispectra of CMB Temperature Anisotropy}

Harmonic coefficients $a_{lm}$ of CMB temperature anisotropies
$\Delta T/T$ are defined as
\begin{equation}
\label{eq:alm}
a_{lm}\equiv \int d\hat{\mathbf n}
\frac{\Delta T}{T}(\hat{\mathbf n})Y_{lm}^\ast(\hat{\mathbf n}).
\end{equation}
Introducing the radiative transfer function $g_{Tl}^{\zeta}$
(or $g_{Tl}^{\cal S}$), they are related to $\zeta$ (or ${\cal S}$) as
\begin{eqnarray}
a_{lm}&=&a_{lm}^\zeta+a_{lm}^{\cal S}, \\
a_{lm}^\zeta &=&4\pi(-i)^l\int\frac{d^3{\mathbf k}}{(2\pi)^3}
\zeta_{\mathbf k} g^\zeta_{Tl}(k) Y_{lm}^\ast(\hat{\mathbf k}), \\
a_{lm}^{\cal S} &=&4\pi(-i)^l\int\frac{d^3{\mathbf k}}{(2\pi)^3}
{\cal S}_{\mathbf k} g^{\cal S}_{Tl}(k) Y_{lm}^\ast(\hat{\mathbf k}).
\end{eqnarray}
The angular power spectra of adiabatic and isocurvature components are
\begin{eqnarray}
\label{eq:cl_aa}
C_l^{\zeta\zeta}&=& \frac{2}{\pi}\int k^2dk
P_{\zeta\zeta}(k)g_{Tl}^\zeta(k)^2, \\
\label{eq:cl_ii1}
C_l^{\cal SS}&=& \frac{2}{\pi}\int k^2dk
P_{\cal SS}(k)g_{Tl}^{\cal S}(k)^2.
\end{eqnarray}

\begin{figure}
\begin{center}
\includegraphics[width=8.5cm]{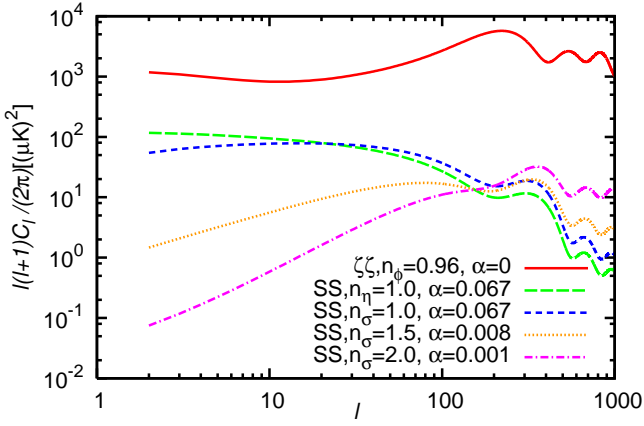}
\caption{Angular power spectra $C_l$ for adiabatic $\zeta\zeta$ and
isocurvature perturbations ${\cal SS}$. The plotted adiabatic
perturbation has the spectral index $n_\phi=0.96$ (solid).  For
isocurvature perturbations, the spectral index of a Gaussian variable
$n_\eta$ is 1 (long-dashed) in the Linear Model and $n_\sigma$ is 1
(short-dashed), 1.5 (dotted) and 2 (dot-dashed) in the Quadratic Model
(see \S \ref{sec:quadratic}). The isocurvature fraction $\alpha$ is
set to be 0.067 ($n_\eta=1$ and $n_\sigma=1$), 0.008 ($n_\sigma=1.5$),
and 0.001 ($n_\sigma=2$) defined at $k_0=0.002$Mpc$^{-1}$.}
\label{fig:cl}
\end{center}
\end{figure}

Fig. \ref{fig:cl} shows the angular power spectrum 
from adiabatic and isocurvature perturbations.
The radiation transfer functions for adiabatic and isocurvature
perturbations are computed using the publicly-available CMBFAST code
\citep{SZ1996}.  The spectral index of isocurvature perturbations in
the Linear Model is $n_\eta=1$. We set $\alpha=0.067$
for $n_\eta=1$ at $k_0=0.002{\rm Mpc}^{-1}$, which is a $2\sigma$ upper
limit from the WMAP 5-year paper \citep{Komatsu2008}.

The total angular bispectrum of CMB is written as the sum of
bispectra with  different combinations of adiabatic and
isocurvature components:
\begin{equation}
b_{l_1l_2l_3}=\sum_{IJK}b_{l_1l_2l_3}^{IJK},
\end{equation}
where each component is defined as
\begin{equation}
\langle a_{l_1m_1}^Ia_{l_2m_2}^Ja_{l_3m_3}^K\rangle
\equiv {\cal G}_{l_1l_2l_3}^{m_1m_2m_3} b_{l_1l_2l_3}^{IJK},
\end{equation}
\begin{eqnarray}
\label{eq:gfac}
{\cal G}_{l_1l_2l_3}^{m_1m_2m_3}
&\equiv &
\int d\hat{\mathbf n}~
Y_{l_1m_1}(\hat{\mathbf n})Y_{l_2m_2}(\hat{\mathbf
n})Y_{l_3m_3}(\hat{\mathbf n}) \\
&=&
\sqrt\frac{(2l_1+1)(2l_2+1)(2l_3+1)}{4\pi}
\left(
\begin{array}{ccc}
l_1 & l_2 & l_3 \\
0 & 0 & 0
\end{array}
\right) \nonumber \\
&&\times
\left(
\begin{array}{ccc}
l_1 & l_2 & l_3 \\
m_1 & m_2 & m_3
\end{array}
\right),
\end{eqnarray}
where $I$ and $J$ denote $\zeta$ or ${\cal S}$. Then,
$b_{l_1l_2l_3}^{IJK}$ can be written as
\begin{eqnarray}
\lefteqn{b_{l_1l_2l_3}^{IJK} = \frac{8}{\pi^3} \int r^2dr \int k_1^2dk_1
\int k_2^2dk_2 \int k_3^2dk_3\,\,
g_{Tl_1}^I(k_1) j_{l_1}(k_1r)}  \nonumber \\
&& \times
g_{Tl_2}^J(k_2) j_{l_2}(k_2r) g_{Tl_3}^K(k_3) j_{l_3}(k_3r) B_{IJK}(k_1,k_2,k_3).
\end{eqnarray}
The explicit form of adiabatic, isocurvature bispectra and their 
cross-correlations are given in Appendix \ref{sec:app}.

Fig. \ref{fig:bis_equi1} shows each component of the CMB bispectrum for
equilateral triangles ($l_1=l_2=l_3=l$). The purely adiabatic
component $b^{\zeta\zeta\zeta}_{lll}$ with $f_{\rm NL}=50$ is plotted
with a solid line.  A long-dashed line represents a mixed component
$b^{(\zeta\zeta{\cal S})}_{lll}$ with $|\cos\theta_{\phi\eta}|=1$ and
the curvaton-type upper limit $\alpha=0.0037$.  The other mixed
component $b^{(\zeta{\cal SS})}_{lll}$ with $|\cos\theta_{\phi\eta}|=0.1$ and
the axion-type limit $\alpha=0.067$ is plotted with
a short-dashed line.  The pure isocurvature component $b^{\cal
SSS}_{lll}$ with the axion-type limit $\alpha=0.067$ is plotted with
dotted lines.  The non-linearity of isocurvature perturbation $f_{\rm
NL}^{\rm (ISO)}$ is set to be $10^4$. The spectral index of
isocurvature perturbation is $1$.

We numerically estimate how large $f_{\rm NL}^{\rm (ISO)}$,
the non-Gaussianity from isocurvature perturbations, should be to obtain
the same values of the signal-to-noise ratio of the CMB bispectrum
as the one derived from the non-Gaussianity from purely adiabatic
perturbations characterized by $f_{\rm NL}$. The signal to noise ratio
is defined as
\begin{equation}
\label{eq:sn}
\left(\frac{S}{N}\right)^2=\sum_{2\le l_1\le l_2\le l_3}I_{l_1l_2l_3}^2
\frac{(b_{l_1l_2l_3}^{IJK})^2}{{\cal C}_{l_1}{\cal C}_{l_2}{\cal C}_{l_3}
\Delta_{l_1l_2l_3}},
\end{equation}
where ${\cal C}_l$ is $C_l$ of purely adiabatic perturbations ($\alpha=0$)
including noise and $I_{l_1l_2l_3}$ is defined as
\begin{equation}
I_{l_1l_2l_3}\equiv
\sqrt{\frac{(2l_1+1)(2l_2+1)(2l_3+1)}{4\pi}}
\left(
\begin{array}{ccc}l_1&l_2&l_3\\0&0&0\end{array}
\right).
\label{eq:I_def}
\end{equation}
The factor $\Delta_{l_1l_2l_3}$ is equal to 6 ($l_1=l_2=l_3$), 2
($l_1=l_2$ or $l_2=l_3$), and 1 ($l_1\ne l_2\ne l_3$). WMAP beam
window functions are included both in the bispectrum and in the power
spectra. The homogeneous noise distribution for the WMAP 5-year data is
used to estimate the noise.

The pure isocurvature term $b^{\cal SSS}$ with $n_\eta=1$ has
non-Gaussianity corresponding to
\begin{equation}
\label{eq:fnl_iii1}
f_{\rm NL}\simeq  15
\left(\frac{\alpha}{0.067}\right)^2
\left(\frac{f_{\rm NL}^{\rm (ISO)}}{10^4}\right),
\end{equation}
where the above equations are valid when $\alpha \ll 1$.
It is found that isocurvature non-Gaussianity reaches $f_{\rm
NL}\sim 10$ only if the non-linearity in isocurvature perturbations
$f_{\rm NL}^{\rm (ISO)}$ is of order $10^4$ because 
isocurvature non-Gaussianity in the Linear Model is suppressed by $\alpha^2$.

When there is a weak correlation between $\phi$ and $\eta$
($|\cos\theta_{\phi\eta}|\ll 1$), $b^{(\zeta{\cal SS})}$ has
non-Gaussianity corresponding to
\begin{equation}
\label{eq:fnl_aii1}
f_{\rm NL}\simeq 9.7\left(\frac{\alpha}{0.067}\right)^{3/2}
\left(\frac{|\cos\theta_{\phi\eta}|}{0.1}\right)
\left(\frac{f_{\rm NL}^{\rm (ISO)}}{10^4}\right).
\end{equation}
If the correlation is strong $|\cos\theta_{\phi\eta}\simeq 1|$
(e.g., curvaton-type isocurvature perturbation), the other correlated
term $b^{(\zeta\zeta{\cal S})}$ becomes important with
a more severe limit $\alpha<0.0037$ for curvaton-type isocurvature
perturbations;
\begin{equation}
\label{eq:fnl_aai1}
f_{\rm NL}\simeq 7.2\left(\frac{\alpha}{0.0037}\right)
\cos^2\theta_{\phi\eta} \left(\frac{f_{\rm NL}^{\rm
(ISO)}}{10^4}\right).
\end{equation}
In both cases, however, isocurvature perturbations need to have a strong
non-linearity $f_{\rm NL}^{(ISO)}\sim 10^4$ in order to generate
non-Gaussianity corresponding to $f_{\rm NL}\sim 10$.  Our result is
consistent with previous theoretical estimations
\citep{KNSST2008,LVW2008}.

\begin{figure}
\includegraphics[width=8.5cm]{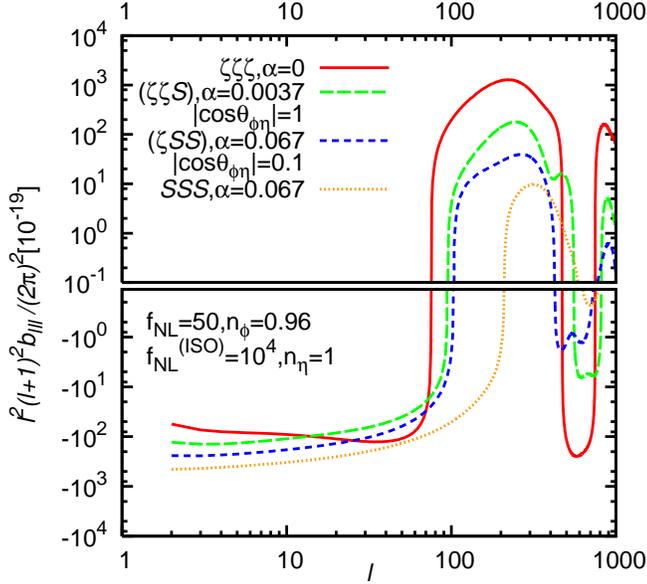}
\begin{center}
\caption{CMB angular bispectra of equilateral configurations
$l^2(l+1)^2b_{lll}/(2\pi)^2$ in the Linear Model; the adiabatic component
$b_{lll}^{\zeta\zeta\zeta}$ (solid), the mixed components
$b_{lll}^{(\zeta\zeta{\cal S})}$ (long-dashed) and
$b_{lll}^{(\zeta{\cal SS})}$ (short-dashed) and the isocurvature component
$b_{lll}^{\cal SSS}$ (dotted).  Upper (Lower) panel shows the positive
(negative) side of bispectra plotted in logarithmic scale. The
adiabatic perturbation has the power-law index $n_{\phi}=0.96$ and its
quadratic amplitude $f_{\rm NL}=50$. The isocurvature perturbation has
$n_{\eta}=1$ and $f_{\rm NL}^{(ISO)}=10^4$.  The fraction of the
isocurvature power spectrum $\alpha$ is set to be an axion-type upper
limit $0.067$ for $b_{lll}^{{\cal SSS}}$ and $b_{lll}^{(\zeta{\cal
SS})}$ with a weak correlation $|\cos\theta_{\phi\eta}|=0.1$ and a
curvaton-type upper limit $0.0037$ for $b_{lll}^{(\zeta\zeta{\cal
S})}$ with a strong correlation $|\cos\theta_{\phi\eta}|=1$.}
\label{fig:bis_equi1}
\end{center}
\end{figure}

\section{Non-Gaussianity of Isocurvature Perturbations II:  Quadratic Model}
\label{sec:quadratic}

\subsection{Initial Perturbation}
In the Quadratic Model (eq.[\ref{eq:iso2}]), the auto and cross power
spectra of $\zeta$ and ${\cal S}$ become
\begin{eqnarray}
\label{eq:pk_ii}
P_{\cal SS}(k)&=&2\int_{L^{-1}}
\frac{d^3{\mathbf p}}{(2\pi)^3}
P_{\sigma\sigma}(p)P_{\sigma\sigma}(|{\mathbf k}+{\mathbf p}|), \\
\label{eq:pk_ai}
P_{\zeta{\cal S}}(k)&=&\frac{6}{5}f_{\rm NL}\int_{L^{-1}}
\frac{d^3{\mathbf p}}{(2\pi)^3}
P_{\phi\sigma}(p)P_{\phi\sigma}(|{\mathbf k}+{\mathbf p}|),
\end{eqnarray}
where a finite box-size $L$ gives an infrared cutoff.  To avoid
assumptions at scales far beyond the present horizon $H_0^{-1}$, $L$
should be set not too much bigger than $H_0^{-1}$ \citep{Lyth2007}.
Hereafter we set $L=30$Gpc.

Using the power-law form as in the equations (\ref{eq:plaw}),
the isocurvature power spectra are written as
\begin{equation}
\label{eq:pkiso2}
\Delta_{\cal SS}(k)=
A_{\sigma}^2F(n_\sigma)\left(\frac{k}{k_0}\right)^{2(n_\sigma-1)},
\end{equation}
where the factor $F$ is approximately given by
\begin{equation}
\label{eq:afac}
F(n_\sigma)=\left\{
\begin{array}{l}
\displaystyle
4\log(kL)~~~~(n_\sigma= 1), \\
\displaystyle
4(n_\sigma-1)^{-1}(1-(kL)^{1-n_\sigma})~~~~(n_\sigma\neq 1),
\end{array}
\right.
\end{equation}
As we take $L$ to be much larger than the range of scales 
we are interested in, the $k$ dependence in $F$ is weak.

As ${\cal S}$ is the square of a Gaussian variable, the
cross-correlation coefficient becomes nearly zero
regardless of $\cos\theta_{\phi\sigma}$;
\begin{eqnarray}
\frac{\Delta_{\zeta{\cal S}}}{(\Delta_{\zeta\zeta}\Delta_{\cal SS})^{1/2}}
& \simeq & f_{\rm NL}(A_\phi F)^{1/2} \cos^2\theta_{\phi\sigma}
\nonumber \\
\label{eq:pcorr_iso2}
& < & f_{\rm NL}(A_\phi F)^{1/2} \le {\cal O}(10^{-2}).
\end{eqnarray}

Following the definition of $\alpha$ in the equation (\ref{eq:alpha}),
the ratio between $A_\phi$ and $A_\sigma$ becomes
\begin{equation}
\label{eq:aratio2}
\frac{A_\sigma}{A_\phi}=\left(\frac{\alpha}{1-\alpha}\right)^{1/2}
(A_\phi F)^{-1/2}.
\end{equation}
The amplitude of $\sigma$ has an additional factor
$(A_\phi F)^{-1/2}\sim {\cal O}(10^{4})$ relative to the amplitude of $\phi$
because the isocurvature perturbation is given as quadratic in $\sigma$
(eq. [\ref{eq:iso2}]).

The bispectra from isocurvature perturbations are written as
\begin{eqnarray}
\label{eq:pbis_iii2}
B_{\cal SSS}(k_1,k_2,k_3)&=&
\frac{8}{3}\int_{L^{-1}}\frac{d^3{\mathbf
p}}{(2\pi)^3}P_{\sigma\sigma}(p) \nonumber \\
&&\times[P_{\sigma\sigma}(|{\mathbf k_1}+{\mathbf p}|)
P_{\sigma\sigma}(|{\mathbf k_2}-{\mathbf p}|) \nonumber \\
&&+P_{\sigma\sigma}(|{\mathbf k_2}+{\mathbf p}|)
P_{\sigma\sigma}(|{\mathbf k_3}-{\mathbf p}|) \nonumber \\
&&+P_{\sigma\sigma}(|{\mathbf k_3}+{\mathbf p}|)
P_{\sigma\sigma}(|{\mathbf k_1}-{\mathbf p}|)].
\end{eqnarray}
When we extract the dominant contributions around the poles, the equation
(\ref{eq:pbis_iii2}) is approximately written as \citep{KNSST2008}
\begin{eqnarray}
\label{eq:bisapp}
B_{\cal SSS}(k_1,k_2,k_3)&\simeq &2\Delta_{\sigma\sigma}(k_b)F(n_\sigma)
[P_{\sigma\sigma}(k_1)P_{\sigma\sigma}(k_2) \nonumber \\
&&+P_{\sigma\sigma}(k_2)P_{\sigma\sigma}(k_3)
+P_{\sigma\sigma}(k_3)P_{\sigma\sigma}(k_1)],
 \nonumber \\ &&
\end{eqnarray}
where $k_b=$Min$(k_1,k_2,k_3)$.

The cross-correlation terms are
\begin{eqnarray}
\label{eq:pbis_aai2}
B_{(\zeta\zeta{\cal S})}(k_1,k_2,k_3)
&=& 2[P_{\phi\sigma}(k_1)P_{\phi\sigma}(k_2)+
\nonumber \\ &&
P_{\phi\sigma}(k_2)P_{\phi\sigma}(k_3)
+P_{\phi\sigma}(k_3)P_{\phi\sigma}(k_1)],
\nonumber \\ &&
\end{eqnarray}
\begin{eqnarray}
\label{eq:pbis_aii2}
B_{(\zeta{\cal SS})}(k_1,k_2,k_3)&=&
\frac{24}{5}f_{\rm NL}\int_{L^{-1}}\frac{d^3{\mathbf
p}}{(2\pi)^3}P_{\sigma\sigma}({\mathbf p}) \nonumber \\
&&\times[P_{\phi\sigma}(|{\mathbf k_1}+{\mathbf p}|)
P_{\phi\sigma}(|{\mathbf k_2}-{\mathbf p}|) \nonumber \\
&&+P_{\phi\sigma}(|{\mathbf k_2}+{\mathbf p}|)
P_{\phi\sigma}(|{\mathbf k_3}-{\mathbf p}|) \nonumber \\
&&+P_{\phi\sigma}(|{\mathbf k_3}+{\mathbf p}|)
P_{\phi\sigma}(|{\mathbf k_1}-{\mathbf p}|)].
\end{eqnarray}

The amplitude of each bispectrum component relative to
the pure adiabatic one at $k_0$ is estimated as
\begin{eqnarray}
\label{eq:bamp_iii2}
\frac{B_{\cal SSS}}{B_{\zeta\zeta\zeta}}&= &
\frac{5}{3}f_{\rm NL}^{-1}\left(\frac{\alpha}{1-\alpha}\right)^{3/2}
(A_\phi F)^{-1/2}, \\
\label{eq:bamp_aai2}
\frac{B_{(\zeta\zeta {\cal S})}}{B_{\zeta\zeta\zeta}}&= &
\frac{5}{3}f_{\rm NL}^{-1}\left(\frac{\alpha}{1-\alpha}\right)^{1/2}
\cos^2\theta_{\phi\sigma} (A_\phi F)^{-1/2}, \\
\label{eq:bamp_aii2}
\frac{B_{(\zeta{\cal SS})}}{B_{\zeta\zeta\zeta}}&= &
3\left(\frac{\alpha}{1-\alpha}\right)
\cos^2\theta_{\phi\sigma}.
\end{eqnarray}
The bispectra $B_{(\zeta\zeta{\cal S})}$ and $B_{\cal SSS}$ has an
additional factor $(A_\phi F)^{-1/2}\sim {\cal O}(10^4)$ relative to
$B_{\zeta\zeta\zeta}$. This is because the isocurvature perturbation
has no Gaussian part (eq.[\ref{eq:iso2}]) and thus its cubic term does
not vanish unlike the adiabatic case.  The isocurvature
non-Gaussianity therefore can be substantial even if $\alpha \sim
0.067$.  We neglect the last term $B_{(\zeta{\cal SS})}$ due to the
current observational limit on $\alpha < 0.067$.

\subsection{Bispectra for CMB Temperature Anisotropy}

The angular power spectra for isocurvature perturbations in the Quadratic
Model are plotted in Fig. \ref{fig:cl} for different spectral indices
$n_\sigma$=1, 1.5 and 2.  We set $\alpha=0.067$ for $n_\sigma=1$
as in the Linear Model. The power spectrum in the Quadratic Model is
slightly different from that in the Linear Model due to the weak scale
dependence of $F$ (eq.[\ref{eq:afac}]).  For other spectral indexes, we set
$\alpha$=0.008 ($n_\sigma=1.5$) and 0.001 ($n_\sigma=2$) so that
the amplitude of $C_l$ around $l$=200 becomes roughly the same.

The angular bispectrum from the isocurvature term $B_{\cal SSS}$ is
approximately given by putting $k_b$ as one of the wavenumbers of
$P_{\sigma\sigma}$ in the equation (\ref{eq:bisapp});
\begin{eqnarray}
\label{eq:cbis_iii2}
b_{l_1l_2l_3}^{\cal SSS}
&=&2 \int r^2dr
[b^{{\cal S},{\cal SS}}_{L l_1}(r)b^{{\cal S},\sigma\sigma}_{L l_2}(r)
b^{{\cal S}}_{NL l_3}(r)
\nonumber \\ &&
+b^{{\cal S},\sigma\sigma}_{L l_1}(r)b^{{\cal S}}_{NL l_2}(r)
b^{{\cal S},{\cal SS}}_{L l_3}(r)
\nonumber \\ &&
+b^{{\cal S}}_{NL l_1}(r)b^{{\cal S},{\cal SS}}_{L l_2}(r)
b^{{\cal S},\sigma\sigma}_{L l_3}(r)],
\end{eqnarray}
where
\begin{equation}
b^{{\cal S},{\cal SS}}_{L l}(r)\equiv\frac{2}{\pi}\int_{L^{-1}} k^2dk
P_{\cal SS}(k)g_{Tl}^{\cal S}(k)j_l(kr).
\end{equation}
In order to test the validity of the above approximation, we compare
the equation (\ref{eq:cbis_iii2}) with full calculation without the
pole approximation which is used to derive the equation (\ref{eq:bisapp}).
The angular bispectrum $B_{\cal SSS}$ without the pole approximation
is analytically given as \citep[eq. (C.7) of ][]{Komatsu2002}
\begin{eqnarray}
\label{eq:cbis_full}
b_{l_1l_2l_3}^{\cal SSS}&=&
\frac{8}{3}\int r^2dr\int_{L^{-1}}
\frac{p^2dp}{(2\pi)^3}
P_{\sigma\sigma}(p)\left[\sum_{l_1^\prime l_2^\prime l}{\cal
F}_{l_2^\prime l_1^\prime l}^{l_1l_2l_3}\right. \nonumber \\
&\times &
\frac{2}{\pi}\int_{L^{-1}} k_1^2dk_1\tilde{P}_{\sigma\sigma~l}^{(+)}(k_1,p)
g_{Tl_1}^{\cal S}(k_1)j_{l^\prime_1}(k_1r)(-i)^{l_1-l_1^\prime} \nonumber \\
&\times&\frac{2}{\pi}\int_{L^{-1}} k_2^2dk_2\tilde{P}_{\sigma\sigma~l}^{(-)}(k_2,p)
g_{Tl_2}^{\cal S}(k_2)j_{l^\prime_2}(k_2r)(-i)^{l_2-l_2^\prime} \nonumber \\
&\times &
\frac{2}{\pi}\int_{L^{-1}} k_3^2dk_3 g_{Tl_3}^{\cal S}(k_3)j_{l_3}(k_3r)
+ ({\rm cyc.})],
\end{eqnarray}
where
\begin{eqnarray}
{\cal F}_{l_2^\prime l_1^\prime l}^{l_1l_2l_3} &\equiv &
\frac{(2l_1^\prime+1)(2l_2^\prime+1)(2l+1)}{4\pi}
\left(\begin{array}{ccc}l_1&l_2&l_3\\0&0&0\end{array}\right)^{-1}
\nonumber \\ &\times &
\left\{\begin{array}{ccc}l_1&l_2&l_3\\l_2^\prime&l_1^\prime&l\end{array}
\right\}
\left(\begin{array}{ccc}l_1^\prime&l_2^\prime&l_3\\0&0&0\end{array}\right)
\left(\begin{array}{ccc}l_1&l_1^\prime&l\\0&0&0\end{array}\right)
\nonumber \\ &\times &
\left(\begin{array}{ccc}l_2&l_2^\prime&l\\0&0&0\end{array}\right)
(-1)^{l_1^\prime +l_2^\prime + l},
\end{eqnarray}
and
\begin{equation}
P_{\sigma\sigma}(|{\mathbf k}\pm{\mathbf p}|)
=\sum_{lm}\tilde{P}_{\sigma\sigma~l}^{(\pm)}
(k,p)Y_{lm}(\hat{\mathbf k})Y_{lm}^\ast(\hat{\mathbf p}).
\end{equation}
When the power spectrum is given in power-law form
as in equation (\ref{eq:plaw}),
\begin{eqnarray}
\frac{k^3\tilde{P}_{\sigma\sigma~l}^{(-)}(k,p)}{2\pi^2}&=&\left\{
\begin{array}{l}
\displaystyle
4\pi A_\sigma\left(\frac{k}{p}\right)^3
\left(\frac{p}{k_0}\right)^{n_\sigma-1}h_l\left(\frac{k}{p}\right),
\\
~~~~~~~~~~~~~~~~~~~~~~~ {\rm when}~p>k+L^{-1}, \\
\\
\displaystyle
4\pi A_\sigma
\left(\frac{k}{k_0}\right)^{n_\sigma-1}h_l\left(\frac{p}{k}\right),
\\
~~~~~~~~~~~~~~~~~~~~~~~ {\rm when}~p<k-L^{-1},
\end{array}
\right. \\
\tilde{P}_{\sigma\sigma~l}^{(+)}(k,p)&=&(-1)^l\tilde{P}_{\sigma\sigma~l}^{(-)}(k,p),
\end{eqnarray}
where $h_l(x)$ is the expansion coefficient by
Legendre function $P_l(z)$ as
\begin{equation}
h_l(y)=\int_{-1}^1 dz(y^2-2yz+1)^{(n_\sigma-4)/2}P_l(z).
\end{equation}
When $n_\sigma$ is equal to $1$, $h_l(y)$ is analytically given as
$2y^l/(1-y^2)$.

\begin{figure*}
\includegraphics[width=8.5cm]{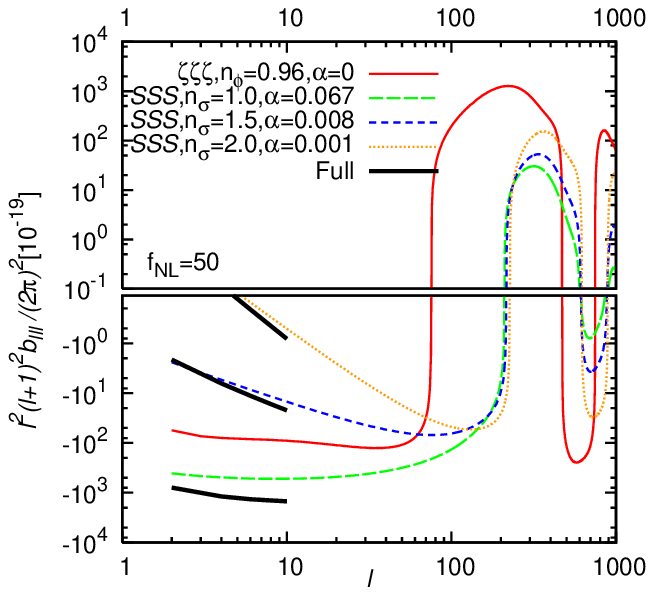}
\includegraphics[width=8.5cm]{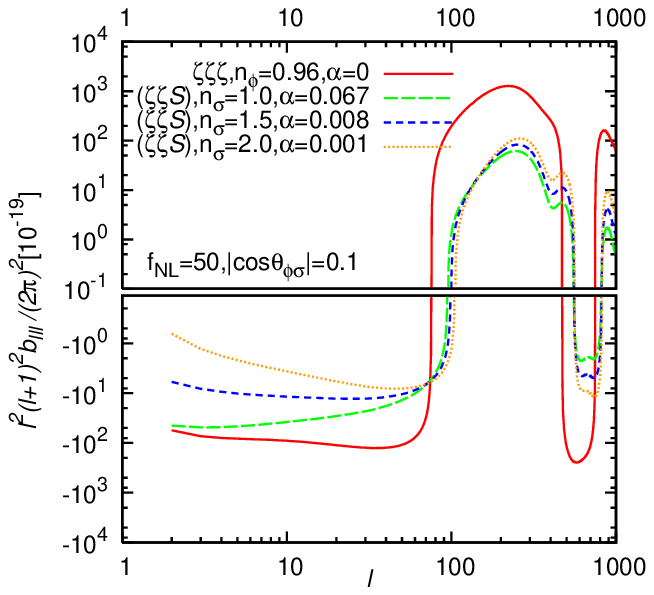}
\begin{center}
\caption{CMB angular bispectra of equilateral configurations
$l^2(l+1)^2b_{lll}/(2\pi)^2$ in the Quadratic Model; the isocurvature
components $b_{lll}^{\cal SSS}$ ({\it Left}) and the mixed components
$b_{lll}^{(\zeta\zeta{\cal S})}$ ({\it Right}). The power-law index of
isocurvature perturbation $n_\sigma$ is set to be 1 (long-dashed), 1.5
(short-dashed) and 2 (dotted). For comparison, the adiabatic
bispectrum $b_{lll}^{\zeta\zeta\zeta}$ with $f_{\rm NL}=50$ is plotted
in both panels (thin solid lines).  The fraction of isocurvature power
spectrum $\alpha$ is 0.067 ($n_\sigma=1$), 0.008 ($n_\sigma=1.5$) and
0.001 ($n_\sigma=2$) defined at $k_0=0.002{\rm Mpc}^{-1}$.  Thick
solid lines in Left panels show the full calculations of isocurvature
terms (eq.[\ref{eq:cbis_full}]) for each $n_\sigma$ at $l\le 10$.  The
box-size $L_{\rm max}$ is set to be 30Gpc.}
\label{fig:bis_equi2}
\end{center}
\end{figure*}

In the left panel of Fig. \ref{fig:bis_equi2}, we compare the
bispectrum of the adiabatic component with those of the isocurvature
components with different spectral indexes $n_\sigma$ of 1, 1.5 and 2.
We set $f_{\rm NL}=50$ and $\alpha=0.067$ for $n_\sigma=1$,
$\alpha=0.008$ for $n_\sigma=1.5$ and $\alpha=0.001$ for $n_\sigma=2$
(same as Fig. \ref{fig:cl}). The box-size $L$ is set to be $30$Gpc.

Computationally, it is very difficult to evaluate the full expression
at high $\ell$. Thus we check the validity of the pole approximations
at low $\ell$, less than $10$.  The thick lines represent the full
calculations of bispectra given in equation (\ref{eq:cbis_full}).
It is found that the isocurvature bispectra approximated as the
equation (\ref{eq:cbis_iii2}) roughly agree with the full calculations
within a factor 2 at least for $l\le 10$. The full calculation has
comparable or larger amplitude for all $n_{\sigma}$, and a slightly
steeper slope than the pole approximation. This indicates that the
isocurvature bispectrum at higher $l$ would become larger than the
pole approximation and then the proper signal may be larger than the
pole approximation.  The amplitude of the isocurvature bispectrum is
proportional to $\alpha^{3/2}$ and therefore the effect on the
estimation of $\alpha$ is suppressed at the power of two-third.

The isocurvature bispectrum in the Quadratic Model depends on the
assumed box-size $L$, which is set to be 30Gpc in this analysis.  The
equation (\ref{eq:bamp_iii2}) indicates that the ratio of the
isocurvature component in the primordial perturbation at
$k=k_0=0.002$Mpc$^{-1}$ is proportional to a box-size dependent factor
$F^{-1/2}$ (eq. [\ref{eq:afac}]).  When $L$ is set to be ten times
larger (300Gpc), the amplitude decreases by 20\% for $n_\sigma=1$, 5\%
for $n_\sigma=1.5$, and 1\% for $n_\sigma=2$.  We find that the overall
amplitude of the CMB bispectrum also decreases at the same level at
$l>10$, while additional large-scale power at more than 30Gpc slightly
increases the amplitude of the bispectrum at smaller $l$ .  The
isocurvature bispectrum depends on $\alpha^{3/2}$ and therefore the
error in $\alpha$ increases $(1/0.8)^{2/3}\simeq$ 16\% for
$n_\sigma=1$.

Using the equation (\ref{eq:sn}), we estimate the isocurvature
non-Gaussianity in terms of $f_{\rm NL}$;
\begin{equation}
\label{eq:fnl_iii2}
f_{\rm NL}=30 \left(\frac{\alpha}{0.067}\right)^{3/2}.
\end{equation}
It is found that the isocurvature non-Gaussianity in the Quadratic Model
can reach $f_{\rm NL}\sim 30$ given the current $2\sigma$
limit on $\alpha$.

The angular bispectrum from the correlation term
$B_{(\zeta\zeta{\cal S})}$ is given as
\begin{eqnarray}
\label{eq:cbis_aai2}
b_{l_1l_2l_3}^{(\zeta\zeta{\cal S})}
&=&2\int r^2dr
[b^{\zeta,\phi\sigma}_{L l_1}(r)b^{\zeta,\phi\sigma}_{L l_2}(r)
b^{{\cal S}}_{NL l_3}(r)
\nonumber \\ &&
+b^{\zeta,\phi\sigma}_{L l_1}(r)b^{{\cal S}}_{NL l_2}(r)
b^{\zeta,\phi\sigma}_{L l_3}(r)
\nonumber \\ &&
+b^{{\cal S}}_{NL l_1}(r)b^{\zeta,\phi\sigma}_{L l_2}(r)
b^{\zeta,\phi\sigma}_{L l_3}(r)].
\end{eqnarray}
The correlated term $b_{l_1l_2l_3}^{(\zeta\zeta{\cal S})}$
is plotted in the right panel of Fig. \ref{fig:bis_equi2}
for different spectral indexes $n_\sigma$=1, 1.5 and 2.
The correlated coefficient $\cos\theta_{\phi\sigma}$ is set to be 0.1.
Using equation (\ref{eq:sn}), the isocurvature non-Gaussianity
with $n_\sigma=1$ corresponds to
\begin{equation}
f_{\rm NL}=240 \left(\frac{\alpha}{0.067}\right)^{1/2}
\cos^2\theta_{\phi\sigma}.
\end{equation}
If $\phi$ and $\sigma$ are strongly correlated initially, the
correlation term can generate substantial non-Gaussianity, while the
effect of the correlation on $C_l$ is negligible as shown in 
equation (\ref{eq:pcorr_iso2}).

\section{Perturbative Formulae of Minkowski Functionals}
\label{sec:mf}

We adopt the perturbation formulae for Minkowski Functionals developed
by \citet{Matsubara2003} to describe the non-Gaussianity of the CMB
temperature anisotropy \citep{HKM2006,Hikage2008}.  We separate the
analytical formulae of Minkowski Functionals into the amplitude and
a function of $\nu$, which is defined as $\Delta T/T$ divided by its
standard deviation, as follows;
\begin{equation}
\label{eq:mf}
V_k(\nu)=A_k v_k(\nu).
\end{equation}
The amplitude $A_k$ is given using the angular power spectrum $C_l$ as
\begin{equation}
\label{eq:mfamp}
A_k=\frac1{(2\pi)^{(k+1)/2}}\frac{\omega_2}{\omega_{2-k}\omega_k}
\left(\frac{\sigma_1}{\sqrt{2}\sigma_0}\right)^k,
\end{equation}
\begin{equation}
\label{eq:var}
\sigma_j^2\equiv \frac1{4\pi}\sum_l(2l+1)\left[l(l+1)\right]^j C_l W^2_l,
\end{equation}
where $\omega_k\equiv \pi^{k/2}/{\Gamma(k/2+1)}$ gives $\omega_0=1$,
$\omega_1=2$, $\omega_2=\pi$ and $W_l$ represents the smoothing kernel
determined by the pixel and beam window functions and an additional
smoothing.  In our analysis, we add a Gaussian kernel
$W_l=\exp[-l(l+1)\theta_s^2/2]$ where $\theta_s$ is a smoothing scale.
For weakly non-Gaussian fields, the function $v_k(\nu)$ can be divided
into the Gaussian term $v_k^{(G)}$ and the non-Gaussian term
$\Delta v_k$;
\begin{eqnarray}
v_k(\nu) &=& v_k^{(G)}(\nu)+\Delta v_k(\nu,f_{\rm NL}), \\
v_k^{(G)} &=& e^{-\nu^2/2}H_{k-1}(\nu), \\
\label{eq:delmf_pb}
 \Delta v_k(\nu,f_{\rm NL}) & = &
e^{-\nu^2/2}\left\{\left[
\frac16S^{(0)}H_{k+2}(\nu)+\frac{k}3S^{(1)}H_k(\nu)
\right.\right. \nonumber \\ & + &
\left.\left.\frac{k(k-1)}6S^{(2)}H_{k-2}(\nu)\right]\sigma_0 + {\cal
O}(\sigma_0^2)\right\},
\end{eqnarray}
where $H_n(\nu)$ is the $n$-th Hermite polynomials.  The non-Gaussian
term $\Delta v_k$ is characterized by three skewness parameters
$S^{(k)}$ at lowest order in $\sigma_0$.  The skewness parameters
$S^{(k)}$ are given by the integral of the reduced bispectrum as
\begin{eqnarray}
S^{(0)}
&=&
\frac3{2\pi\sigma_0^4}\sum_{2\le l_1\le l_2\le l_3}
I^2_{l_1l_2l_3}b_{l_1l_2l_3}W_{l_1}W_{l_2}W_{l_3}, \label{eq:s0_cmb}\\
S^{(1)}
&=&
\nonumber
\frac3{8\pi\sigma_0^2\sigma_1^2}\sum_{2\le l_1\le l_2\le l_3}
\left[l_1(l_1+1)+l_2(l_2+1)+l_3(l_3+1)\right] \\
& &\times
I^2_{l_1l_2l_3}b_{l_1l_2l_3}W_{l_1}W_{l_2}W_{l_3}, \label{eq:s1_cmb}\\
S^{(2)}
&=&
\nonumber
\frac3{4\pi\sigma_1^4}\sum_{2\le l_1\le l_2\le l_3}
\left\{\left[l_1(l_1+1)+l_2(l_2+1)-l_3(l_3+1)\right]\right. \\
& &
\left.\times l_3(l_3+1)+({\rm cyc.})\right\}
I^2_{l_1l_2l_3}b_{l_1l_2l_3}W_{l_1}W_{l_2}W_{l_3}, \label{eq:s2_cmb}
\end{eqnarray}
where $I_{l_1l_2l_3}$ was previously defined in Eq.~(\ref{eq:I_def}).

\begin{figure*}
\begin{center}
\includegraphics[width=5.8cm]{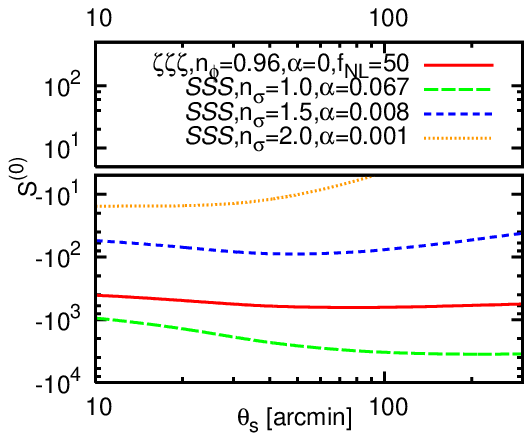}
\includegraphics[width=5.8cm]{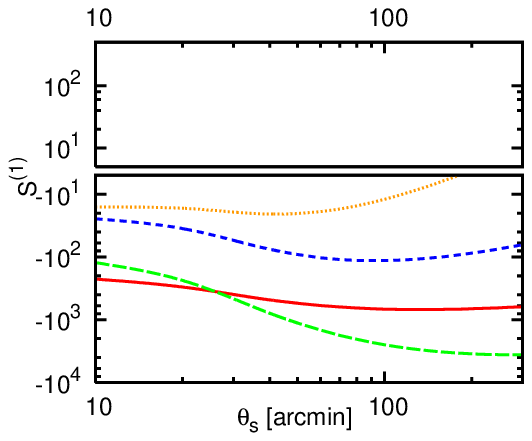}
\includegraphics[width=5.8cm]{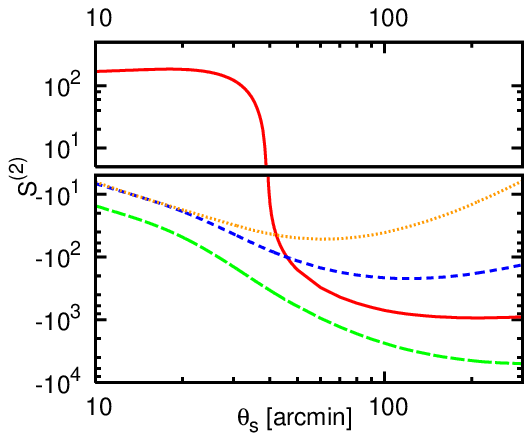}
\includegraphics[width=5.8cm]{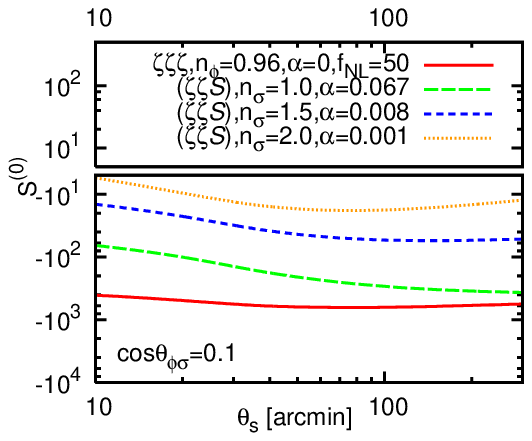}
\includegraphics[width=5.8cm]{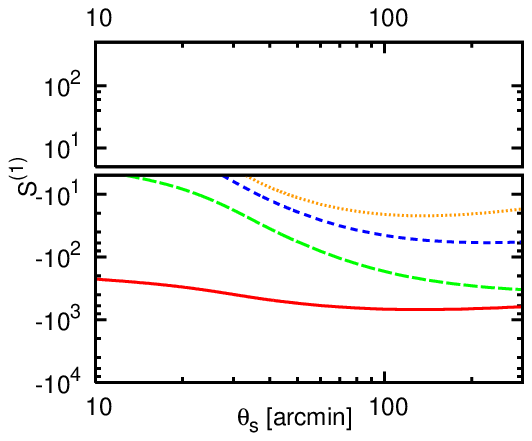}
\includegraphics[width=5.8cm]{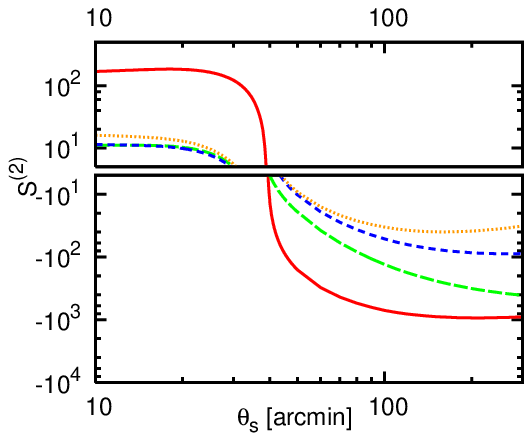}
\caption{Skewness parameters $S^{(k)}$ ({\it left:} $k=0$, {\it
center:} $k=1$, {\it right:} $k=2$) of each component in the Quadratic
Model plotted as a function of $\theta_s$.  The skewness values for a pure
isocurvature component are plotted in upper panels, while those for a
mixed component are plotted in lower panels. The parameters of
isocurvature perturbation are $n_\sigma$=1 with $\alpha=0.067$
(long-dashed), $n_\sigma$=1.5 with $\alpha=0.008$ (short-dashed),
and $n_\sigma$=2 with $\alpha=0.001$ (dotted). For comparison, the
adiabatic skewness with $n_\phi=0.96$ and $f_{\rm NL}=50$ are plotted
with solid lines in all panels.}
\label{fig:skew_iso}
\end{center}
\end{figure*}

In Fig. \ref{fig:skew_iso}, we plot the three skewness parameters
$S^{(k)}$ from each component of bispectrum in the Quadratic Model as a
function of $\theta_s$. Upper panels show the skewness of isocurvature
components with different spectral indices $n_\sigma=$1, 1.5 and 2 in
comparison with the adiabatic case. The skewness from a mixed
component $b^{(\zeta\zeta{\cal S})}$ is plotted in the lower panels. 
The smaller scale (higher $l$) information in the bispectrum is reflected in
skewness parameters with a higher number (that is $S^{(2)}$ rather than
$S^{(0)}$) because it is weighted towards higher $l$.  We can
extract further detailed scale-dependent information by changing
the smoothing scale $\theta_s$.

\begin{figure*}
\begin{center}
\includegraphics[width=5.8cm]{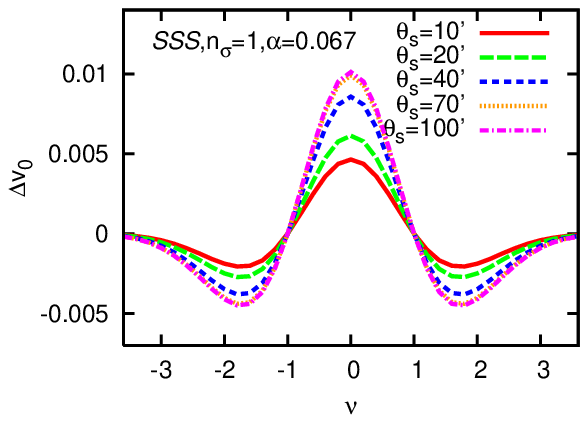}
\includegraphics[width=5.8cm]{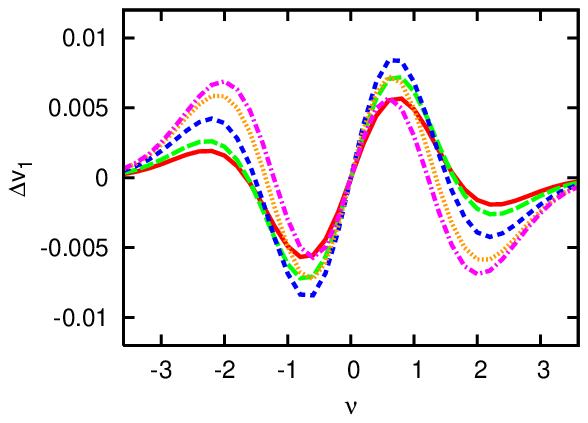}
\includegraphics[width=5.8cm]{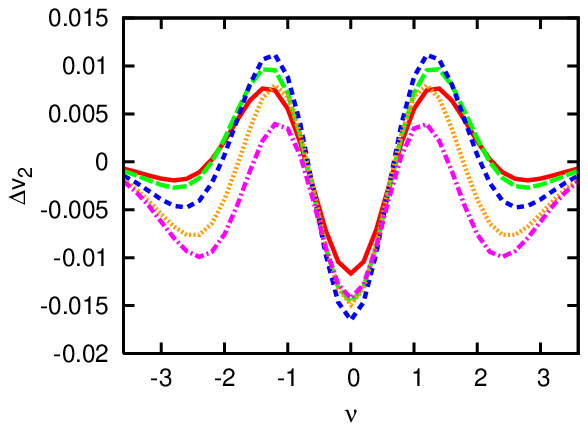}
\caption{Non-Gaussian term of Minkowski Functionals $\Delta v_k$
(eq.[\ref{eq:delmf_pb}]) from isocurvature bispectrum in Quadratic
Model ({\it left:} $k=0$, {\it center:} $k=1$, {\it right:} $k=2$).
The isocurvature perturbation has a spectral index $n_\sigma$=1 and
the fraction $\alpha$ is $0.067$.  The smoothing scales are shown by 
different lines: $\theta_s=10$ (solid), $\theta_s=20$
(long-dashed), $\theta_s=40$ (short-dashed) $\theta_s=70$ (dotted),
and $\theta_s=100$ (dot-dashed) arcmin.}
\label{fig:mf}
\end{center}
\end{figure*}

Fig. \ref{fig:mf} illustrates an example of non-Gaussian effects on
each Minkowski Functional $\Delta v_k$ (eq.[\ref{eq:delmf_pb}]) at
different smoothing scales $\theta_s$=10, 20, 40, 70 and 100 arcmin.
Here we consider isocurvature non-Gaussianity in the Quadratic Model with
$n_{\sigma}=1$ and $\alpha=0.067$. It is found that the
non-Gaussian effect on Minkowski Functionals is 1\% or less.

\section{Limits on Isocurvature Non-Gaussianity from WMAP Data}
\label{sec:wmaplimit}

We use the WMAP 5-year data to constrain the non-Gaussianity associated with
primordial isocurvature perturbations. We use the linearly co-added
maps for Q, V and W frequency bands with $N_{\rm side}=512$.  The
co-added maps are masked with the {\it Kq75} galaxy mask including
the point-source mask provided by \cite{WMAPfg}, which leaves $71.8\%$ of
the sky available for the data analysis. The field is smoothed with a
Gaussian filter at a scale of $\theta_s$. We obtain the normalized
Minkowski Functional (eq.[\ref{eq:delmf_pb}]) for the WMAP data using the
same procedure described in \citet{Hikage2008}.

We use Bayes's theorem to find a probability of
$\alpha$ or other combined parameters, together denoted by $x$,
from the observed set of $\Delta \nu^{\rm (obs)}$ as follows;
\begin{equation}
{\cal P}(x|\Delta \nu^{\rm (obs)})\propto{\cal L}(\Delta v^{\rm (obs)}|x)
{\cal P}(x),
\end{equation}
where ${\cal L}$ is the likelihood function of $\Delta v^{\rm (obs)}$
when a non-Gaussian parameter has a value $x$ and the probability
${\cal P}(x)$ represents the prior for $x$. In general, we need to
analyze all non-Gaussian components together, however we consider the
cases in which a single non-Gaussian term dominates over the other 
terms for simplicity; in the Linear Model, the pure isocurvature term
(eq.[\ref{eq:cbis_iii1}]) dominates when there is no correlation
between $\phi$ and $\eta$. If there exists a weak correlation, the
correlation term of $(\zeta{\cal SS})$ (eq. [\ref{eq:cbis_aii1}])
becomes important. The other correlation term of $(\zeta\zeta{\cal
S})$ (eq. [\ref{eq:cbis_aai1}]) dominates when the correlation is
strong ($\cos\theta_{\phi\eta}\simeq 1$). In these three cases, 
we set the parameter $x$ to be $\alpha^2 f_{\rm
NL}^{\rm (ISO)}$, $\alpha^{3/2}\cos\theta_{\phi\eta}f_{\rm NL}^{\rm
(ISO)}$ and $\alpha\cos^2\theta_{\phi\eta}f_{\rm NL}^{\rm (ISO)}$. 
In the Quadratic Model, the pure isocurvature term (eq.[\ref{eq:cbis_iii2}])
dominates at $\cos\theta_{\phi\sigma}\ll 1$.  Then we set $x$ to be
$\alpha$ (not $\alpha^{3/2}$).  When the correlation between $\phi$
and $\sigma$ is strong, the correlation term (eq.[\ref{eq:cbis_aai2}])
is dominant and then $x$ is $\alpha^{1/2}\cos^2\theta_{\phi\sigma}$. 
We assume a flat prior for all $x$.  We furthermore impose a non-negative 
constraint on the parameters in the Quadratic Model, $\alpha$ and
$\alpha^{1/2}\cos^2\theta_{\phi\sigma}$ by definition of $\alpha$
(eq.[\ref{eq:alpha}]).

The likelihood function is computed by
\begin{eqnarray}
\label{eq:covmatrix}
  -2\ln{\cal L}(\Delta v^{\rm (obs)}|x)&\propto
  &\sum_{ij}
  [\Delta v_i^{\rm (obs)}-\Delta v_i^{\rm (theory)}(x)]
  C^{-1}_{ij} \nonumber \\
  & \times &
  [\Delta v_j^{\rm (obs)}-\Delta v_j^{\rm
(theory)}(x)],
\end{eqnarray}
where $i$ and $j$ denote the binning number of threshold values $\nu$,
different kinds of Minkowski Functional $k$, and smoothing scales
parameterized with $\theta_s$. We choose 18 threshold values at an
equal spacing in the range of $\nu$ from $-3.6$ to $3.6$.  The full
covariance matrix $C_{ij}$ is estimated from 1000 Gaussian simulation
maps from purely adiabatic perturbations. They include the pixel and
beam window function, {\it Kq75} survey mask, and inhomogeneous noise
for WMAP 5-year maps. Applying our procedure to limit the
non-Gaussianity from curvature perturbations by putting $x$ as $f_{\rm
NL}$, we obtain $-63<f_{\rm NL}<76$ at 95\% CL (Hikage et al. in
preparation).

Table \ref{tab:fiso_lin} lists the mean and one-sigma error of the
isocurvature non-Gaussianity in the Linear Model characterized by
$\alpha\cos^2\theta_{\phi\eta} f_{\rm NL}^{\rm (ISO)}$,
$\alpha^{3/2}\cos\theta_{\phi\eta} f_{\rm NL}^{\rm (ISO)}$, and
$\alpha^2f_{\rm NL}^{\rm (ISO)}$. Significant isocurvature
non-Gaussian signals are not found.  If isocurvature perturbations
exist and there is no correlation between $\phi$ and $\eta$,
the non-linear parameter $f_{\rm NL}^{\rm (ISO)}$ is constrained
from isocurvature non-Gaussianity for a fixed $\alpha$ to be 
\begin{equation}
f_{\rm NL}^{\rm (ISO)} = (-3300\pm 13000)(\alpha/0.067)^{-2}.
\end{equation}
If there is a strong correlation between $\phi$ and $\eta$
represented by curvaton-type isocurvature perturbations
($\cos\theta_{\phi\eta}=-1$), the correlated term $(\zeta\zeta{\cal
S})$ becomes important and then its non-Gaussianity is limited as
\begin{equation}
f_{\rm NL}^{\rm (ISO)} = (4900\pm 43000)(\alpha/0.0037)^{-1}.
\end{equation}

Table \ref{tab:fiso_alpha} lists the maximum likelihood value
$\alpha_{\rm ML}$, at which ${\cal P}(\alpha)$ has a maximum value,
and the 95\% confidence limit of $\alpha$ when the pure isocurvature
term in the Quadratic Model dominates. We do not find a significant
non-zero value of $\alpha$. The upper limit of $\alpha$ is given by
\begin{eqnarray}
\alpha &<& 0.070 \quad (n_\sigma=1),\nonumber\\
\alpha &<& 0.042 \quad (n_\sigma=1.5),\nonumber\\
\alpha &<& 0.0064 \quad (n_\sigma=2),
\end{eqnarray}
at 95\% CL.  Our constraint on $\alpha$ for $n_\sigma=1$ is comparable
to that from the joint limit from WMAP(TT and TE spectra)+BAO+SN,
which is smaller than 0.067 (95\% CL), obtained by
\citet{Komatsu2008}.  The limits on the correlated term
$\alpha^{1/2}\cos^2\theta_{\phi\sigma}$ are also listed in Table
\ref{tab:fais}.  Our result roughly agrees with \citet{BL2006}
who showed the fraction of a Gaussian-squared
component of primordial curvature perturbation is limited to be less
than $0.18$, which corresponds to $\alpha<0.031$, when $|f_{\rm
NL}|<100$.  The difference from their analysis is that we calculate a
Gaussian-squared component of CMB isocurvature bispectra including its
full radiative transfer function.

\section{Implications for axion isocurvature}
\label{sec:implications}

Here we briefly discuss implications of our results in some explicit
models. Although, for the linear model, the constraint from
non-Gaussianity obtained here is not as strong as that from the power
spectrum, the constraint for the quadratic model is severe as we
showed in the previous section.  Thus we consider
the case of the axion as discussed in \citet{KNSST2008}.  Assuming that
Pecci-Quinn symmetry has already been spontaneously broken during
inflation, the mean value of the axion field can be written as
\begin{equation}
\label{eq:axion}
a  = f_a \theta_a,
\end{equation}
where $f_a$ is the axion decay constant and $\theta_a$ is the phase of
the axion.  During inflation, the axion field has quantum
fluctuations 
\begin{equation}
\label{eq:delta_theta}
\delta a = \frac{H_{\rm inf}}{2\pi},
\end{equation}
where $H_{\rm inf}$ is the Hubble parameter during inflation.  When
the average value of the axion is much less the mean-square
inhomogeneity of $a$, i.e., when $f_a \theta_a \le H_{\rm inf} /
2\pi$, the density fluctuation of $a$ is written as
\begin{equation}
\label{eq:rho_a}
\frac{\delta \rho_a}{\rho_a}  = \left( \frac{\delta a}{a} \right)^2.
\end{equation}
Thus power spectrum for isocurvature fluctuations can be written as
\begin{equation}
\label{eq:P_a}
\frac{k^3 P_{SS}}{2\pi^2} = \frac{(\Omega_ah^2)^2}{(\Omega_{\rm cdm}h^2)^2},
\end{equation}
where $\Omega_a$ is the energy density of the axion at present.
When $ f_a \theta < H_{\rm inf} / 2\pi$, the abundance of the
axion is given by  \citep{Turner1986}
\begin{equation}
\label{eq:omega_a}
\Omega_a h^2
=
0.2 \left( \frac{f_a}{10^{12} ~{\rm GeV}} \right)^{-0.825}
\left( \frac{H_{\rm inf}/2 \pi }{10^{12} ~{\rm GeV}} \right)^{2}.
\end{equation}
Since the total amplitude of primordial perturbations should be
$\Delta_{\rm tot}\simeq \Delta_{\zeta\zeta}+\Delta_{\cal SS}
=k^3/2\pi^2 (P_{\zeta\zeta}+P_{\cal SS})= 2.4 \times 10^{-9}$ which is
required from WMAP5, by using the constraint on $\alpha$
presented in the previous section, $\alpha < 0.070$, we obtain a
limit for the Hubble parameter during inflation
\begin{equation}
H_{\rm inf} <  1.7 \times 10^{10}~
\left( \frac{f_a}{10^{12} ~{\rm GeV}} \right)^{0.41}~{\rm GeV},
\end{equation}
where we adopt $\Omega_{\rm cdm} h^2 = 0.108$ which is the mean value
for a $\Lambda$CDM model from the WMAP5 analysis. The condition
$f_a \theta_a < H_{\rm inf}/2 \pi$ gives
\begin{equation}
\theta_a < 2.7 \times 10^{-3} \left( \frac{f_a}{10^{12}
~{\rm GeV}} \right)^{-0.59}.
\end{equation}

Although our results obtained in this paper may also have implications
for other models with isocurvature fluctuations, a detailed study of
this issue will be given elsewhere.

\section{Summary and Conclusions}
\label{sec:summary}

We have explored the effect of non-Gaussianity from primordial isocurvature
perturbations on CMB temperature anisotropies. Considering the linear
and quadratic forms of isocurvature perturbations, which are applicable
to a wide range of theoretical models, we derived theoretical
expressions for bispectra and Minkowski Functionals of CMB temperature
maps with isocurvature non-Gaussianity. We find that the amplitude of
a quadratic correction $f_{\rm NL}^{(ISO)}$ in the Linear Model (a
Gaussian variable plus its quadratic correction) needs to be of the order
of $10^4$ to generate CMB non-Gaussianity at a level of $f_{\rm
NL}\sim 10$. The isocurvature non-Gaussianity in the Quadratic Model 
(quadratic in a Gaussian variable without linear terms) can reach
$f_{\rm NL}=30$ while respecting the current upper limit on the 
isocurvature contribution to the power spectrum $\alpha<0.067$.  Isocurvature
perturbations provide a possible mechanism to explain primordial
non-Gaussianity recently suggested from the observed bispectrum of the CMB
\citep{YW2008,Komatsu2008}.

We give limits on isocurvature non-Gaussianity from Minkowski
Functionals for the WMAP 5-year data. In the Quadratic Model of isocurvature
perturbations, we obtain a stringent limit $\alpha<0.070$ (95\% CL)
from the non-Gaussianity, which is comparable to the
current constraints from WMAP $TT$ and $TE$ spectra, BAO and SN
combined $\alpha<0.067$ (95\% CL). We apply our results
to a QCD axion isocurvature model and then obtain a limit for the
Hubble parameter and the phase of the axion.

We estimate isocurvature non-Gaussianity in the Quadratic Model
using the pole approximation (eq.[\ref{eq:bisapp}]). The validity can
be checked by comparing Minkowski Functionals with non-Gaussian
simulations with a Gaussian-squared perturbation.  We plan to perform
this analysis in the near future.

We employ Minkowski Functionals to characterize non-Gaussianity in CMB
maps. The application of our work to other higher-order statistics
is important to utilize non-Gaussian
information in a more complete manner.  Different statistics are
sensitive to different aspects of density fields and they are
affected by possible observational systematics (e.g., foregrounds and
point sources) in a different way. There is actually some friction
between the limits on $f_{\rm NL}$ from Minkowski Functionals and the 
bispectrum \citep{Hikage2008}; Minkowski Functionals analysis indicate
a maximum likelihood value around nearly 0, while bispectrum analysis
favours a more positive $f_{\rm NL}$ around 50 or more.  Complementary
analyses with different statistical approaches will provide a more
robust way to analyze primordial non-Gaussianity.

\section*{Acknowledgments}
We appreciate Toyokazu Sekiguchi and Fuminobu Takahashi for kindly
providing their data to check our calculations. We also appreciate
Eiichiro Komatsu and Misao Sasaki for useful advice and
discussions. We thank David Wands for careful reading of the paper and
for the useful comments.  We thank the organizers of the workshop
``Non-Gaussianity from Fundamental Physics'' at DAMTP, Cambridge, 8-10
September 2008 for their kind invitation.  C.~H. acknowledges support
from the Particle Physics and Astronomy Research Council grant number
PP/C501692/1 and a JSPS (Japan Society for the Promotion of Science)
fellowship.  K.K. is supported by ERC, RCUK and STFC. This work is
also supported in part by the Sumitomo Foundation (T.T.), and by
Grant-in-Aid for Scientific Research from the Ministry of Education,
Science, Sports, and Culture of Japan No.\,19740145 (T.T.),
No.\,18740157, and No.\,19340054 (M.Y.).



\begin{table*}
\caption{Mean values and 1 $\sigma$ uncertainties of each isocurvature
non-Gaussianity parameter in the Linear Model; the pure isocurvature term
$\alpha^2f_{\rm NL}^{\rm (ISO)}$ (eq.[\ref{eq:cbis_iii1}]) and the
correlated terms with adiabatic perturbations
$\alpha^{3/2}\cos\theta_{\phi\eta} f_{\rm NL}^{\rm (ISO)}$
(eq.[\ref{eq:cbis_aii1}]) and $\alpha\cos^2\theta_{\phi\eta} f_{\rm
NL}^{\rm (ISO)}$ (eq.[\ref{eq:cbis_aai1}]). The limits are obtained
from Minkowski Functionals for WMAP 5-year data at different smoothing
scales $\theta_s$ and their combination.}
\begin{center}
\begin{tabular}{cccc}
  \hline\hline
$\theta_s$ [arcmin] &
$\alpha^2f_{\rm NL}^{\rm (ISO)}$ &
$\alpha^{3/2}\cos\theta_{\phi\eta}f_{\rm NL}^{\rm (ISO)}$ &
$\alpha\cos^2\theta_{\phi\eta}f_{\rm NL}^{\rm (ISO)}$
\\ \hline
           100 & $~33 \pm 84$ & $~\ 26 \pm 62$ & $~91 \pm 220$\\
            70 & $~36 \pm 74$ & $~\ 28 \pm 56$ & $104 \pm 210$\\
            40 & $~71 \pm 78$ & $~\ 56 \pm 61$ & $256 \pm 230$\\
            20 & $~\ 0 \pm 89$ & $ -3 \pm 66$ & $~73 \pm 240$\\
            10 & $~\ 1 \pm 91$ & $-30 \pm 60$ & $~41 \pm 230$\\
10,~20,~40,~70,~100 		
               & $-15 \pm 60$ & $-18 \pm 43$ & $~18 \pm 160$\\ \hline
\end{tabular}
\end{center}
\label{tab:fiso_lin}
\end{table*}

\begin{table*}
\caption{Limits on $\alpha$ for the Quadratic Model from Minkowski
Functionals for WMAP 5-year data at different smoothing scales
$\theta_s$. The listed values are maximum likelihood values
$\alpha_{\rm ML}$ and 95\% CL on $\alpha$. We neglect non-Gaussianity
from curvature perturbations and their cross-correlation. We impose
a non-negative condition on $\alpha$ following from its 
definition in equation
(\ref{eq:alpha}).}
\begin{center}
\begin{tabular}{ccccccc}
  \hline\hline
& \multicolumn{2}{c}{$n_\sigma=1$}
& \multicolumn{2}{c}{$n_\sigma=1.5$}
& \multicolumn{2}{c}{$n_\sigma=2$}
\\
\cline{2-7}
\raisebox{1.5ex}[0pt]{$\theta_s$ [arcmin]} &
\raisebox{-1ex}{$\alpha_{\rm ML}$} &
\raisebox{-1ex}{95\% CL} &
\raisebox{-1ex}{$\alpha_{\rm ML}$} &
\raisebox{-1ex}{95\% CL} &
\raisebox{-1ex}{$\alpha_{\rm ML}$} &
\raisebox{-1ex}{95\% CL}
\\ \hline
           100 & 0.037 & $<0.115$ & 0.022 & $<0.100$ & 0.005 & $<0.047~$\\
            70 & 0.036 & $<0.105$ & 0.029 & $<0.079$ & 0.013 & $<0.030~$\\
            40 & 0.051 & $<0.112$ & 0.010 & $<0.057$ & 0     & $<0.014~$\\
            20 & 0     & $<0.101$ & 0     & $<0.048$ & 0     & $<0.0085$\\
            10 & 0     & $<0.093$ & 0     & $<0.047$ & 0     & $<0.0071$\\
10,~20,~40,~70,~100 & 0 & $<0.070$ & 0    & $<0.042$ & 0     & $<0.0064$\\
\hline
\end{tabular}
\end{center}
\label{tab:fiso_alpha}
\end{table*}

\begin{table*}
\caption{Same as Table \ref{tab:fiso_alpha} but for the parameter
$\alpha^{1/2}\cos^2\theta_{\phi\sigma}$ where the correlation term
$b^{(\zeta\zeta{\cal S})}_{l_1l_2l_3}$ in the Quadratic Model 
dominates. We also impose a non-negative condition on
$\alpha^{1/2}\cos^2\theta_{\phi\sigma}$.}
\begin{center}
\begin{tabular}{ccccccc}
  \hline\hline
& \multicolumn{2}{c}{$n_\sigma=1$}
& \multicolumn{2}{c}{$n_\sigma=1.5$}
& \multicolumn{2}{c}{$n_\sigma=2$}
\\
\cline{2-7}
\raisebox{1.5ex}[0pt]{$\theta_s$ [arcmin]} &
\raisebox{-1ex}{$(\alpha^{1/2}\cos^2\theta_{\phi\sigma})_{\rm ML}$} &
\raisebox{-1ex}{95\% CL} &
\raisebox{-1ex}{$(\alpha^{1/2}\cos^2\theta_{\phi\sigma})_{\rm ML}$} &
\raisebox{-1ex}{95\% CL} &
\raisebox{
-1ex}{$(\alpha^{1/2}\cos^2\theta_{\phi\sigma})_{\rm ML}$} &
\raisebox{-1ex}{95\% CL}
\\ \hline
       100 &  0.018 &  $<0.095$ &  0.018 &  $<0.12~$ &  0.010~ & $<0.089$ \\
        70 &  0.021 &  $<0.092$ &  0.021 &  $<0.11~$ &  0.014~ & $<0.073$ \\
        40 &  0.045 &  $<0.12~$ &  0.040 &  $<0.14~$ &  0.017~ & $<0.075$ \\
        20 &  0.012 &  $<0.098$ &  0.026 &  $<0.11~$ &  0.013~ & $<0.041$ \\
        10 &  0     &  $<0.071$ &  0.003 &  $<0.063$ &  0.0026 & $<0.019$ \\
10,~20,~40,~70,~100 & 0 &  $<0.049$ & 0  &  $<0.039$ &  0      & $<0.012$ \\
\hline
\end{tabular}
\end{center}
\label{tab:fais}
\end{table*}

\appendix
\section{Adiabatic, Isocruvature, \& Mixed components of CMB
Angular Bispectrum}
\label{sec:app}

Explicit forms of the angular bispectra from pure adiabatic and
isocurvature mode are calculated as \citep{KS2001}
\begin{eqnarray}
\label{eq:cbis_adi}
b_{l_1l_2l_3}^{\zeta\zeta\zeta}&=&\frac{6}{5}f_{\rm NL}\int r^2dr[
b^{\zeta,\phi\phi}_{L l_1}(r)b^{\zeta,\phi\phi}_{L l_2}(r)
b^{\zeta}_{NL l_3}(r)
\nonumber \\ &&
+b^{\zeta,\phi\phi}_{L l_1}(r)b^{\zeta}_{NL l_2}(r)
b^{\zeta,\phi\phi}_{L l_3}(r)
\nonumber \\ &&
+b^{\zeta}_{NL l_1}(r)b^{\zeta,\phi\phi}_{L l_2}(r)
b^{\zeta,\phi\phi}_{L l_3}(r)], \\
\label{eq:cbis_iii1}
b_{l_1l_2l_3}^{\cal SSS}&=&2f_{\rm NL}^{(ISO)}\int r^2dr[
b^{{\cal S},\eta\eta}_{L l_1}(r)b^{{\cal S},\eta\eta}_{L l_2}(r)
b^{{\cal S}}_{NL l_3}(r)
\nonumber \\ &&
+b^{{\cal S},\eta\eta}_{L l_1}(r)b^{{\cal S}}_{NL l_2}(r)
b^{{\cal S},\eta\eta}_{L l_3}(r)
\nonumber \\ &&
+b^{{\cal S}}_{NL l_1}(r)b^{{\cal S},\eta\eta}_{L l_2}(r)
b^{{\cal S},\eta\eta}_{L l_3}(r)],
\end{eqnarray}
where
\begin{eqnarray}
b^{\zeta,\phi\phi}_{L l}(r)&\equiv &
\frac{2}{\pi}\int k^2dkP_{\phi\phi}(k)g_{Tl}^\zeta(k)j_l(kr), \\
b^{{\cal S},\eta\eta}_{L l}(r)&\equiv &\frac{2}{\pi}\int k^2dkP_{\eta\eta}(k)
g_{Tl}^{\cal S}(k)j_l(kr), \\
b^{\zeta}_{NL l}(r)&\equiv &
\frac{2}{\pi}\int k^2dkg_{Tl}^\zeta(k)j_l(kr), \\
b^{{\cal S}}_{NL l}(r)&\equiv &\frac{2}{\pi}\int k^2dk
g_{Tl}^{\cal S}(k)j_l(kr).
\end{eqnarray}

The angular bispectra from the correlation terms are given by
\begin{eqnarray}
\label{eq:cbis_aai1}
b_{l_1l_2l_3}^{(\zeta\zeta{\cal S})}
&\equiv
& b_{l_1l_2l_3}^{\zeta\zeta{\cal S}}+b_{l_1l_2l_3}^{\zeta{\cal S}\zeta}+
b_{l_1l_2l_3}^{{\cal S}\zeta\zeta} \nonumber \\
&=&2 f_{\rm NL}^{(ISO)}\int r^2dr
\left[b^{\zeta,\phi\eta}_{L l_1}(r)b^{\zeta,\phi\eta}_{L l_2}(r)
b^{{\cal S}}_{NL l_3}(r) \right.
\nonumber \\ &&
+b^{\zeta,\phi\eta}_{L l_1}(r)b^{{\cal S}}_{NL l_2}(r)
b^{\zeta,\phi\eta}_{L l_3}(r)
\nonumber \\ && \left.
+b^{{\cal S}}_{NL l_1}(r)b^{\zeta,\phi\eta}_{L l_2}(r)
b^{\zeta,\phi\eta}_{L l_3}(r) \right]
\nonumber \\ &&
+\frac65 f_{\rm NL} \int r^2dr \times
\nonumber \\ &&
\left[b^{\zeta,\phi\phi}_{L l_1}(r)
\left\{b^{{\cal S},\phi\eta}_{L l_2}(r)b^{\zeta}_{NL l_3}(r)
+b^{{\cal S},\phi\eta}_{L l_3}(r)b^{\zeta}_{NL l_2}(r)
\right\} \right.
\nonumber \\ &&
+b^{\zeta,\phi\phi}_{L l_2}(r)
\left\{b^{{\cal S},\phi\eta}_{L l_1}(r)b^{\zeta}_{NL l_3}(r)
+b^{{\cal S},\phi\eta}_{L l_3}(r)b^{\zeta}_{NL l_1}(r)
\right\}
\nonumber \\ && \left.
+b^{\zeta,\phi\phi}_{L l_3}(r)
\left\{b^{{\cal S},\phi\eta}_{L l_1}(r)b^{\zeta}_{NL l_2}(r)
+b^{{\cal S},\phi\eta}_{L l_2}(r)b^{\zeta}_{NL l_1}(r)
\right\}
\right], \nonumber \\ &&
\end{eqnarray}
\begin{eqnarray}
\label{eq:cbis_aii1}
b_{l_1l_2l_3}^{(\zeta{\cal SS})}
&\equiv & b_{l_1l_2l_3}^{\zeta{\cal SS}}+b_{l_1l_2l_3}^{{\cal S}\zeta{\cal S}}+
b_{l_1l_2l_3}^{{\cal SS}\zeta} \nonumber \\
&=&\frac65 f_{\rm NL}\int r^2dr
\left[b^{{\cal S},\phi\eta}_{L l_1}(r)b^{{\cal S},\phi\eta}_{L l_2}(r)
b^{\zeta}_{NL l_3}(r) \right.
\nonumber \\ &&
+b^{{\cal S},\phi\eta}_{L l_1}(r)b^{\zeta}_{NL l_2}(r)
b^{{\cal S},\phi\eta}_{L l_3}(r)
\nonumber \\ && \left.
+b^{\zeta}_{NL l_1}(r)b^{{\cal S},\phi\eta}_{L l_2}(r)
b^{{\cal S},\phi\eta}_{L l_3}(r) \right]
\nonumber \\ &&
+2 f_{\rm NL}^{(ISO)} \int r^2dr \times
\nonumber \\ &&
\left[b^{\zeta,\phi\eta}_{L l_1}(r)
\left\{b^{{\cal S},\eta\eta}_{L l_2}(r)b^{\cal S}_{NL l_3}(r)
+b^{{\cal S},\eta\eta}_{L l_3}(r)b^{\cal S}_{NL l_2}(r)
\right\} \right.
\nonumber \\ &&
+b^{\zeta,\phi\eta}_{L l_2}(r)
\left\{b^{{\cal S},\eta\eta}_{L l_1}(r)b^{\cal S}_{NL l_3}(r)
+b^{{\cal S},\eta\eta}_{L l_3}(r)b^{\cal S}_{NL l_1}(r)
\right\}
\nonumber \\ && \left.
+b^{\zeta,\phi\eta}_{L l_3}(r)
\left\{b^{{\cal S},\eta\eta}_{L l_1}(r)b^{\cal S}_{NL l_2}(r)
+b^{{\cal S},\eta\eta}_{L l_2}(r)b^{\cal S}_{NL l_1}(r)
\right\}
\right], \nonumber \\ &&
\end{eqnarray}
where
\begin{eqnarray}
b^{\zeta,\phi\eta}_{L l}(r) &\equiv&
\frac{2}{\pi}\int k^2dkP_{\phi\eta}(k)
g_{Tl}^\zeta(k)j_l(kr), \\
b^{{\cal S},\phi\eta}_{L l}(r) &\equiv&
\frac{2}{\pi}\int k^2dkP_{\phi\eta}(k)
g_{Tl}^{\cal S}(k)j_l(kr).
\end{eqnarray}
The terms proportional to $f_{\rm NL}$ in
$b_{l_1l_2l_3}^{(\zeta\zeta{\cal S})}$ and $b_{l_1l_2l_3}^{(\zeta{\cal
SS})}$ are very small relative to the adiabatic bispectrum under the
current observational constraints and thus they are neglected in the
following analysis.


\begin{thebibliography}{99}
\bibitem[\protect\citeauthoryear{Alishahiha et al.}{2004}]{AST2004}
Alishahiha,~M., Silverstein,~E., Tong,~D., 2004, Phys. Rev. D, 70, 123505

\bibitem[\protect\citeauthoryear{Arkami-Hamed et al.}{2004}]{ACMZ2004}
Arkani-Hamed,~N., Creminelli,~P., Mukohyama,~S., Zaldarriaga,~M., 2004,
JCAP, 4, 1

\bibitem[\protect\citeauthoryear{Arroja, Mizuno \& Koyama}{2008}]{AMK2008}
Arroja,~F., Mizuno,~S., Koyama,~K., 2008, JCAP, 8, 15

\bibitem[\protect\citeauthoryear{Assadullahi et al.}{2007}]{AVW2007}
Assadullahi,~H., Valiviita,~J., Wands,~D., 2007, Phys.\ Rev.\  D, 76, 103003

\bibitem[\protect\citeauthoryear{Bartolo et al.}{2004}]{BMR2004}
Bartolo,~N., Matarrese,~S., Riotto,~A., 2004, Phys. Rev. D, 69, 043503

\bibitem[\protect\citeauthoryear{Bean et al.}{2006}]{BDP2006}
Bean,~R., Dunkley,~J., Pierpaoli,~E., 2006, Phys. Rev. D, 74, 063503

\bibitem[\protect\citeauthoryear{Beltran}{2008}]{Beltran2008}
Beltran,~M., 2008, Phys. Rev. D, 78, 023530

\bibitem[\protect\citeauthoryear{Boubekeur \& Creminelli}{2006}]{BC2006}
Boubekeur,~L., Creminelli,~P., 2006, Phys.\ Rev.\  D, 73, 103516

\bibitem[\protect\citeauthoryear{Boubekeur \& Lyth}{2006}]{BL2006}
Boubekeur,~L., Lyth,~D.~H., 2006, Phys. Rev. D, 73, 021301

\bibitem[\protect\citeauthoryear{Buchbinder et. al.}{2008}]{BKO2008}
Buchbinder,~E.~I., Khoury,~J., Ovrut,~B.~A., 2008, Phys. Rev. Lett, 100, 171302

\bibitem[\protect\citeauthoryear{Chen, Easther \& Lim}{2007a}]{CEL2007}
Chen,~X., Easther,~R., Lim,~E.~A., 2007, JCAP, 6, 23

\bibitem[\protect\citeauthoryear{Chen et al.}{2007b}]{CHKS2007}
Chen,~X., Huang,~M.~x., Kachru,~S., Shiu,~G., 2007, JCAP, 1, 2

\bibitem[\protect\citeauthoryear{Creminelli et al.}{2007}]{Creminelli2007}
Creminelli,~P., Senatore,~L., Zaldarriaga,~M., Tegmark,~M., 2007, JCAP, 3, 5

\bibitem[\protect\citeauthoryear{Dvali, Gruzinov \& Zaldarriaga}{2004}]{DGZ2004}
Dvali,~G., Gruzinov,~A., Zaldarriaga,~M., 2004, Phys. Rev. D, 69, 083505

\bibitem[\protect\citeauthoryear{Dunkley et al.}{2009}]{Dunkley2008}
Dunkley,~J. et al., 2009, ApJS, 180, 306

\bibitem[\protect\citeauthoryear{Enqvist \& Sloth}{2002}]{ES2002}
Enqvist,~K., Sloth,~M.~S., 2002, Nucl. Phys. B, 626, 395

\bibitem[\protect\citeauthoryear{Enqvist \& Nurmi}{2005}]{EN2005}
Enqvist,~K., Nurmi,~S., 2005, JCAP, 10, 13

\bibitem[\protect\citeauthoryear{Enqvist \& Takahashi}{2008}]{ET2008}
Enqvist,~K., Takahashi,~T., 2008, JCAP 9, 12

\bibitem[\protect\citeauthoryear{Gold et al.}{2009}]{WMAPfg}
Gold,~B. et al., 2009, ApJS, 180, 265

\bibitem[\protect\citeauthoryear{Hikage, Komatsu \& Matsubara}{2006}]{HKM2006}
Hikage,~C., Komatsu,~E., Matsubara,~T., 2006, ApJ, 653, 11

\bibitem[\protect\citeauthoryear{Hikage et al.}{2008}]{Hikage2008}
Hikage,~C., Matsubara,~T., Coles,~P., Liguori,~M., Hansen,~F.~K.,
Matarrese,~S., 2008, MNRAS, 389, 1439

\bibitem[\protect\citeauthoryear{Huang}{2008}]{Huang2008}
Huang,~Q.~G., 2008, Phys. Lett. B, 669, 260

\bibitem[\protect\citeauthoryear{Ichikawa et al.}{2008a}]{ISTY2008a}
Ichikawa,~K., Suyama,~T., Takahashi,~T., Yamaguchi,~M., 2008,
Phys. Rev. D, 78, 023513

\bibitem[\protect\citeauthoryear{Ichikawa et al.}{2008b}]{ISTY2008b}
Ichikawa,~K., Suyama,~T., Takahashi,~T., Yamaguchi,~M., 2008,
Phys.\ Rev.\  D, 78, 063545

\bibitem[\protect\citeauthoryear{Kawasaki \& Sekiguchi}{2007}]{KS2007}
Kawasaki,~M., Sekiguchi,~T., 2008, Prog. Theor. Phys., 120, 995

\bibitem[\protect\citeauthoryear{Kawasaki et al.}{2008}]{KNSST2008}
Kawasaki,~M., Nakayama,~K., Sekiguchi,~T., Suyama,~T., Takahashi,~F.,
2008, JCAP, 11, 19

\bibitem[\protect\citeauthoryear{Kawasaki, Nakayama \& Takahashi}{2009}]{KNT2009}
Kawasaki,~M., Nakayama,~K., Takahashi,~F., 2009, JCAP, 1, 2

\bibitem[\protect\citeauthoryear{Kawasaki et al.}{2009}]{KNSST2009}
Kawasaki,~M., Nakayama,~K., Sekiguchi,~T., Suyama,~T., Takahashi,~F.,
2009, JCAP, 1, 42

\bibitem[\protect\citeauthoryear{Kofman}{2003}]{Kofman2003}
Kofman,~L., preprint (astro-ph/0303614)

\bibitem[\protect\citeauthoryear{Komatsu \& Spergel}{2001}]{KS2001}
Komatsu,~E., Spergel,~D.~N., 2001, Phys. Rev. D, 63, 63002

\bibitem[\protect\citeauthoryear{Komatsu}{2002}]{Komatsu2002}
Komatsu,~E., preprint (astro-ph/0206039)

\bibitem[\protect\citeauthoryear{Komatsu et al.}{2008}]{Komatsu2008}
Komatsu,~E. et al., 2009, ApJS, 180, 330

\bibitem[\protect\citeauthoryear{Koyama et al.}{2007}]{KMVW2007}
Koyama,~K., Mizuno,~S., Vernizzi,~F., Wands,~D., 2007, JCAP, 11, 24

\bibitem[\protect\citeauthoryear{Langlois et al.}{2008a}]{LRST2008}
Langlois,~D., Renaux-Petel,~S., Steer,~D.~A., Tanala,~T., 2008,
Phys. Rev. Lett., 101, 061301

\bibitem[\protect\citeauthoryear{Langlois, Vernizzi \& Wands}{2008b}]{LVW2008}
Langlois,~D., Vernizzi,~F., Wands,~D., 2008, JCAP, 12, 4

\bibitem[\protect\citeauthoryear{Lehners \& Steinhardt}{2008}]{LS2008}
Lehners,~J.~L., Steinhardt,~P.~J., 2008, Phys. Rev. D, 77, 063533

\bibitem[\protect\citeauthoryear{Linde \& Mukhanov}{1997}]{LM1997}
Linde,~A.~D., Mukhanov,~V., 1997, Phys. Rev. D, 56, R535

\bibitem[\protect\citeauthoryear{Lyth, Ungarelli \& Wands}{2003}]{Lyth2003}
Lyth,~D.H., Ungarelli,~C., Wands,~D., 2003,  Phys. Rev. D, 67, 23503

\bibitem[\protect\citeauthoryear{Lyth \& Wands}{2002}]{LW2002}
Lyth,~D.~H., Wands,~D., 2002  Phys.\ Lett.\  B, 524, 5

\bibitem[\protect\citeauthoryear{Lyth}{2007}]{Lyth2007}
Lyth,~D.~H., 2007, JCAP, 12, 16

\bibitem[\protect\citeauthoryear{Matsubara}{2003}]{Matsubara2003}
Matsubara,~T., 2003, ApJ, 584, 1

\bibitem[\protect\citeauthoryear{Malik \& Lyth}{2006}]{ML2006}
Malik,~K.~A., Lyth,~D.~H., 2006, JCAP, 9, 8

\bibitem[\protect\citeauthoryear{Mollerach}{1990}]{Mollerach1990}
Mollerach,~S., 1990, Phys. Rev. D, 42, 313

\bibitem[\protect\citeauthoryear{Moroi \& Takahashi}{2001}]{MT2001}
Moroi,~T., Takahashi,~T., 2001, Phys. Lett. B, 522, 215

\bibitem[\protect\citeauthoryear{Moroi \& Takahashi}{2008}]{MT2008}
Moroi,~T., Takahashi,~T., 2009, Phys. Lett. B, 671, 339

\bibitem[\protect\citeauthoryear{Naruko \& Sasaki}{2009}]{NS2009}
Naruko,~A., Sasaki,~M., 2009, Prog. Theor. Phys., 121, 193

\bibitem[\protect\citeauthoryear{Peebles}{1999}]{Peebles1999}
Peebles,~P.~J.~E., 1999, ApJ, 510, 531

\bibitem[\protect\citeauthoryear{Sasaki}{2008}]{Sasaki2008}
Sasaki,~M., 2008, Prog. Theor. Phys., 120, 159

\bibitem[\protect\citeauthoryear{Sasaki, Valiviita \& Wands}{2006}]{SVW2006}
Sasaki,~M., Valiviita,~J., Wands,~D., 2006, Phys. Rev. D, 74, 103003

\bibitem[\protect\citeauthoryear{Seljak \& Zaldarriaga}{1996}]{SZ1996}
Seljak,~U., Zaldarriaga,~M. 1996, ApJ, 469, 437

\bibitem[\protect\citeauthoryear{Slosar et al.}{2008}]{Slosar2008}
Slosar,~A., Hirata,~C., Seljak,~U., Ho,~S., Padmanabhan,~N.,
2008, JCAP, 8, 31

\bibitem[\protect\citeauthoryear{Suyama \& Takahashi}{2008}]{ST2008}
Suyama,~T., Takahashi,~F., 2008, JCAP, 9, 7

\bibitem[\protect\citeauthoryear{Suyama \& Yamaguchi}{2008}]{SY2008}
Suyama,~T., Yamaguchi,~M., 2008, Phys. Rev. D, 023505

\bibitem[\protect\citeauthoryear{Turner}{1986}]{Turner1986}
Turner,~M.~S., 1986, Phys.\ Rev.\  D, 33, 889

\bibitem[\protect\citeauthoryear{Yadav \& Wandelt}{2008}]{YW2008}
Yadav,~A.~P.~S., Wandelt,~B.,~D., 2008, Phys. Rev. Lett., 100, 181301

\bibitem[\protect\citeauthoryear{Zaldarriaga}{2004}]{Zaldarriaga2004}
Zaldarriaga,~M., 2004,  Phys.\ Rev.\  D,\  69, 043508


\end{thebibliography}
\end{document}